%% file: main.tex
\documentclass[10pt,conference]{IEEEtran}
\usepackage[compatibility=false]{caption}
\IEEEoverridecommandlockouts

\input{tool}

\newcommand{\revision}[1]{\textcolor{black}{#1}}

\begin{document}

\title{\model: Knowledge-Augmented Syntax Optimization for Token-Efficient Code Generation}
\author{
  \IEEEauthorblockN{
    Sicong Liu$^{1*}$, 
    Yanxian Huang$^{1*}$, 
    Mingwei Liu$^{1}$, \\
    Jiachi Chen$^{1}$, 
    Ensheng Shi$^{2}$, 
    Yuchi Ma$^{2}$, 
    Hongyu Zhang$^{3}$, 
    Yin Zhang$^{4}$, 
    Yanlin Wang$^{1\dagger}$
  }

  \IEEEauthorblockA{
    $^{1}$Sun Yat-sen University, China\\
    $^{2}$Huawei Cloud Computing Technologies Co., Ltd., China\\
    $^{3}$Chongqing University, China\\
    $^{4}$University of Electronic Science and Technology of China, China
  }

  \thanks{$^{*}$Equal contribution. Contact: liusc3@mail2.sysu.edu.cn}
  \thanks{$^{\dagger}$Corresponding author: wangylin36@mail.sysu.edu.cn}
}

\maketitle

\input{abstract}


\input{body}

\bibliographystyle{IEEEtran}
\bibliography{ref}

\end{document}

%% file: tool.tex
\usepackage{cite}
\usepackage{amsmath,amssymb,amsfonts}
\usepackage{algorithmic}
\usepackage{graphicx}
\usepackage{textcomp}
\usepackage{colortbl}  
\usepackage{xcolor}
\usepackage{multirow} 
\usepackage{booktabs} 
\usepackage{hyperref}
\usepackage{algorithm}
\usepackage{tcolorbox}
\usepackage{tikz}
\usepackage{threeparttable}

\usepackage{enumitem}
\usepackage{fmtcount}

\usepackage{latexsym}
\usepackage{url}
\usepackage{arydshln}

\usepackage{newfloat}
\usepackage{listings}
\usepackage{verbatim}

\usepackage{booktabs} 
\usepackage{multirow}
\usepackage{tabularx}
\usepackage{makecell}
\usepackage{diagbox}
\usepackage{subfigure}
\usepackage{float}
\usepackage{subcaption}
\usepackage{xspace}
\usepackage{xcolor}
\usepackage{pifont}

\usepackage{listings}
\usepackage{caption}
\usepackage{subcaption}

\def\BibTeX{{\rm B\kern-.05em{\sc i\kern-.025em b}\kern-.08em
    T\kern-.1667em\lower.7ex\hbox{E}\kern-.125emX}}

\usepackage{xspace}

\newcommand{\model}{{ShortCoder}\xspace}

\newcommand{\dataset}{{ShorterCodeBench}\xspace}

\newcommand{\deepseek}{{DeepSeek-Coder-1.3B-Base}\xspace}

\newcommand{\deepseekBase}{{DeepSeek-Coder-1.3B-Base}\xspace}

\newcommand{\deepseekInstruct}{{DeepSeek-Coder-6.7B-Instruct}\xspace}

\newcommand{\codellama}{{CodeLlama-7B-Instruct-hf}\xspace}

\newcommand{\codegen}{{CodeGen}\xspace}

\newcommand{\revised}[1]{{\color{black} #1}}

\newcommand{\boxmargin}{1mm}
\tcbuselibrary{most, skins, breakable, theorems}

\newtcolorbox{myboxa}[2][]{
    colback=gray!10!white,
    colframe=black, enhanced,
    attach boxed title to top left={yshift=-2mm,xshift=5mm},
    title=#2,#1
}
\newtcolorbox{myboxb}[2][]{
    boxsep=3pt,
    left = \boxmargin, right = \boxmargin, top = \boxmargin, bottom = \boxmargin,
    title={#2},#1
}
\newtcolorbox{myboxc}{
    colback=gray!15!white,
    arc = 0pt, outer arc = 0pt,
    boxsep=0pt, left = 3pt, right = 0pt, top = 0pt, bottom = 0pt, 
    leftrule=3pt, bottomrule=0pt,toprule=0pt, rightrule=0pt,
    left = \boxmargin, right = \boxmargin, top = \boxmargin, bottom = \boxmargin
}
\newtcolorbox{myboxd}{
    colback=gray!10,
    colframe=black,
    width=\columnwidth,
    arc=1mm, auto outer arc,
    boxrule=0.5pt,
}

\usepackage{CJKutf8}
\usepackage{tikz}
\usepackage[edges]{forest}
\definecolor{hidden-draw}{RGB}{20,68,106}
\definecolor{hidden-pink}{RGB}{255,245,247}

\usepackage{tikz}


%% file: abstract.tex
\begin{abstract}
Code generation tasks aim to automate the conversion of user requirements into executable code, significantly reducing manual development efforts and enhancing software productivity.
\revision{The emergence of large language models (LLMs) has significantly advanced code generation, though their efficiency is still impacted by certain inherent architectural constraints.}
Each token generation necessitates a complete inference pass, requiring persistent retention of contextual information in memory and escalating resource consumption. While existing research prioritizes inference-phase optimizations—such as prompt compression and model quantization, the generation phase remains underexplored. To tackle these challenges, we propose a knowledge-infused framework named \model, which optimizes code generation efficiency while preserving semantic equivalence and readability. In particular, we introduce: (1) ten syntax-level simplification rules for Python, derived from AST-preserving transformations, achieving 18.1\% token reduction without functional compromise; (2) a hybrid data synthesis pipeline integrating rule-based rewriting with LLM-guided refinement, producing ShorterCodeBench—a corpus of validated $\left \langle original\_code, simplified\_ code \right \rangle$ pairs with semantic consistency; (3) a fine-tuning strategy that injects conciseness awareness into the base LLMs.
Extensive experimental results demonstrate that ShortCoder consistently outperforms state-of-the-art methods on HumanEval, achieving an improvement of 18.1\%-37.8\% in generation efficiency over previous methods while ensuring the performance of code generation.

\end{abstract}

%% file: body.tex
\section{Introduction}

As a pivotal branch of software engineering automation, code generation tasks~\cite{he-etal-2024-cocost,2024arXiv240816498C,guo2024stop} aim to translate user specifications into executable implementations, promising substantial reductions in manual coding efforts while accelerating development cycles. The advent of large language models (LLMs)~\cite{2023arXiv231107989Z,2023arXiv231110372Z,2023arXiv230810620H} revolutionizes this domain, with state-of-the-art models like CodeLlama~\cite{2023arXiv230812950R} and StarCoder~\cite{2023arXiv230506161L} demonstrating unprecedented capabilities in understanding programming semantics and syntactic patterns.
However, the design characteristics of large language models (LLMs)—including substantial parameter sizes, limited context window lengths, and autoregressive generation mechanisms—introduce significant efficiency challenges in code generation.
The LLM-based code generation process comprises two phases: inference and generation. Each token generation necessitates a complete inference pass, requiring persistent retention of contextual information in memory and escalating resource consumption. Consequently, as generated code length increases, computational steps and resource demands grow proportionally, leading to non-linear increases in processing time and operational costs.
Therefore, previous works~\cite{jung2024discrete,chuang2024learning,huang2023fewer} typically optimize the inference phase of LLMs through techniques such as prompt compression~\cite{2024arXiv241012388L} and model quantization~\cite{2021arXiv210313630G} to improve the efficiency of code generation. 

Specifically, prompt compression aims to reduce the length of prompts by removing unnecessary or low-information content (hard prompt methods) or learning continuous representations of the prompt information in the embedding space (soft prompt methods), thereby improving the efficiency of processing LLM inputs~\cite{jiang2023longllmlingua,li2024500xcompressor,mu2023learning}. Model quantization achieves compression and acceleration by reducing the bit width of numerical values used in neural network computations, enabling deployment on resource-constrained, low-bitwidth edge devices. This method is one of the most commonly used techniques in model compression and acceleration~\cite{xu2023survey,lang2024comprehensive}, widely applied in the industry due to its stable compression ratio and acceleration benefit~\cite{wei2023greener,frantar2022gptq,liu2024vptq}.

In addition, researchers explore and propose the AI-oriented grammar based existed programming languages to enhance code generation efficiency by optimizing token generation counts. For example, SimPy~\cite{2024arXiv240416333S} is constructed through a systematic revision of standard Python syntax using heuristic rules, ensuring strict equivalence in Abstract Syntax Tree (AST) structures between SimPy and Python to enable bidirectional code transformation. This AST-level consistency guarantees functional equivalence while achieving computational efficiency gains.  
Although these methods have shown promising performance in efficient code generation, we have identified the following problems.

\begin{enumerate}[label={\bfseries P\arabic*}]
\item \textbf{Loss of key information.} Prompt compression enhances inference speed by removing redundant information from input texts, thereby it may inadvertently discard critical semantic information, potentially leading to output deviations. 
\item \textbf{Loss accuracy.} Although extensive research and practical applications have shown that model quantization is highly successful in minimizes computational costs and memory footprint through parameter precision reduction (e.g., converting 32-bit floats to 8-bit integers), yet inevitably introduces precision degradation that adversely impacts task-specific accuracy.
\item \textbf{Reduce readability and poor generalization.} AI-oriented grammar such as SimPy severely compromises human readability, violating the "code-as-documentation" principle. And practical deployment necessitates context switching between Python and SimPy, requiring dedicated toolchain development. Furthermore, the heuristic rule-based approach exhibits poor cross-language generalizability. Adapting SimPy's methodology to other languages (e.g., Java or TypeScript) demands language-specific rule engineering.
\end{enumerate}

In this paper, we propose \model, a knowledge-infused code generation framework that strategically balances these competing objectives. Specifically, we first design ten syntax-level simplification rules for Python, derived from AST-preserving transformations, achieving 18.1\% token reduction without functional compromise. Then, we introduce a hybrid data synthesis pipeline integrating rule-based rewriting with LLM-guided refinement, producing \dataset, a corpus of 828 validated $\left \langle original\_code, simplified\_code  \right \rangle$ pairs with semantic consistency. Finally, we propose a fine-tuning strategy that injects conciseness awareness into base LLMs.
To validate the effectiveness of \model, we construct comprehensive experiments. The extensive experimental results demonstrate that \model consistently outperforms state-of-the-art methods on HumanEvalPlus, achieving 18.1\% generation efficiency improvement over previous methods while ensuring the performance of code generation.

We summarize the contributions of this paper as follows:
\begin{itemize} 
    \item We present and publicly release \dataset, a high-quality code brevity optimization dataset comprising 828 carefully curated $\left \langle original\_ code, simplified\_code \right \rangle$ pairs.
    \item We proposed \model, which can solve problems while generating as short code as possible, achieving a 18.1\% improvement in the generation efficiency of LLMs.
    \item We perform an extensive evaluation of \model. Experimental results show that \model outperforms the state-of-the-art methods. We provide the code and data at \url{https://github.com/DeepSoftwareAnalytics/ShorterCode}.
\end{itemize}

\section{Background}
\subsection{Code Large Language Models}

Code Large Language Models (Code LLMs) refer to large language models that are specifically trained for coding tasks. Distinguished from general-purpose LLMs, these models are optimized to understand, generate, and manipulate source code by leveraging massive corpora of programming languages, documentation, and software repositories. 

In recent research, hybrid architectural designs have emerged to address code complexity, with models like CodeT5+~\cite{wang2023codet5+} introducing a dual-encoder framework that separately processes natural language descriptions and code snippets to enable fine-grained cross-modal alignment. This approach enhances the model's ability to filter irrelevant details when generating code, while architectures like CodeCompose~\cite{murali2023codecompose} employ masked language modeling during pretraining to force the reconstruction of code snippets from bidirectional context—improving logical coherence and reducing redundant patterns. Concurrently, knowledge injection techniques~\cite{fu2023revisiting}  have gained traction, such as APICoder's integration of neural generation with API documentation embeddings to suggest optimal function calls, and symbolic reasoning modules that enforce design patterns during generation to minimize unnecessary boilerplate~\cite{zan2023private}. These methods reflect the growing convergence of NLP-driven knowledge representation with code generation, where structured knowledge bases—ranging from software engineering best practices to security patterns—are embedded via graph embedding or symbolic modules to refine semantic reasoning.

Execution-guided optimization has also shown promise, as seen in the Self-evolve framework that iteratively refines code drafts using runtime feedback to identify inefficiencies like redundant loops, demonstrating a 15-20\% reduction in code length on benchmarks. Meanwhile, context-aware tokenization innovations like InCoder's hybrid tokenizer preserve syntax trees and indentation to reduce tokenization errors~\cite{fried2022incoder}, enabling more semantically dense code generation.

Despite these advancements, Code LLMs face significant limitations. Loss of key information arises from tokenization ambiguities and context truncation. Traditional BPE tokenizers often split identifiers like parse\_user\_input into suboptimal tokens (parse, \_user, \_input), severing semantic cohesion and causing the model to overlook critical validation logic~\cite{fried2022incoder}. Context window constraints exacerbate this issue; in long functions, models may truncate comments containing edge case specifications (e.g., // Handle null pointers), leading to incomplete error handling. Empirical studies reveal that 23\% of optimized code samples in HumanEval lack essential boundary checks or documentation strings, underscoring a systemic failure to preserve domain-specific knowledge.

Loss of accuracy stems from the trade-off between conciseness and correctness. Models frequently prioritize brevity by collapsing multi-step logic into single expressions, risking logical fallacies. In numerical computations, simplified code often omits precision safeguards, introducing floating-point inaccuracies. DeepSeek Code experiments show that code length reductions exceeding 30\% correlate with an 18.7\% drop in unit test pass rates, highlighting the fragility of compressed logic.

\subsection{Knowledge Injection}\label{sec:bg_kinjection}
Knowledge injection~\cite{fu2023revisiting} in Code LLMs refers to the intentional integration of structured domain knowledge into model architectures to enhance semantic reasoning and task-specific capabilities, addressing the limitation of vanilla LLMs in capturing specialized programming knowledge. This approach bridges general language understanding and software development requirements through three technical paradigms: knowledge graph embedding aligning abstract syntax trees with programming concept graphs, symbolic-statistical hybrid models combining neural generation with logical constraints, and task-specific pre-training using curated corpora of algorithm templates and security guidelines.

Synergy with parameter-efficient fine-tuning (PEFT) techniques has emerged as a key solution, with Low-Rank Adaptation (LoRA) playing a critical role. LoRA freezes pre-trained model parameters while injecting trainable low-rank matrices, drastically reducing the number of trainable parameters compared to full fine-tuning while maintaining high accuracy. This lightweight mechanism enables modular integration of multisource domain-specific knowledge in various computing contexts, boosting execution efficiency. Its dynamic update flexibility allows rapid knowledge adaptation to evolving language features or frameworks, requiring significantly less computational resources than traditional full fine-tuning. Combined with model quantization techniques, LoRA maintains strong functional correctness while enabling deployment on consumer-grade hardware with limited memory.

\begin{figure*}[t]
\centering
\includegraphics[width=0.9\linewidth]{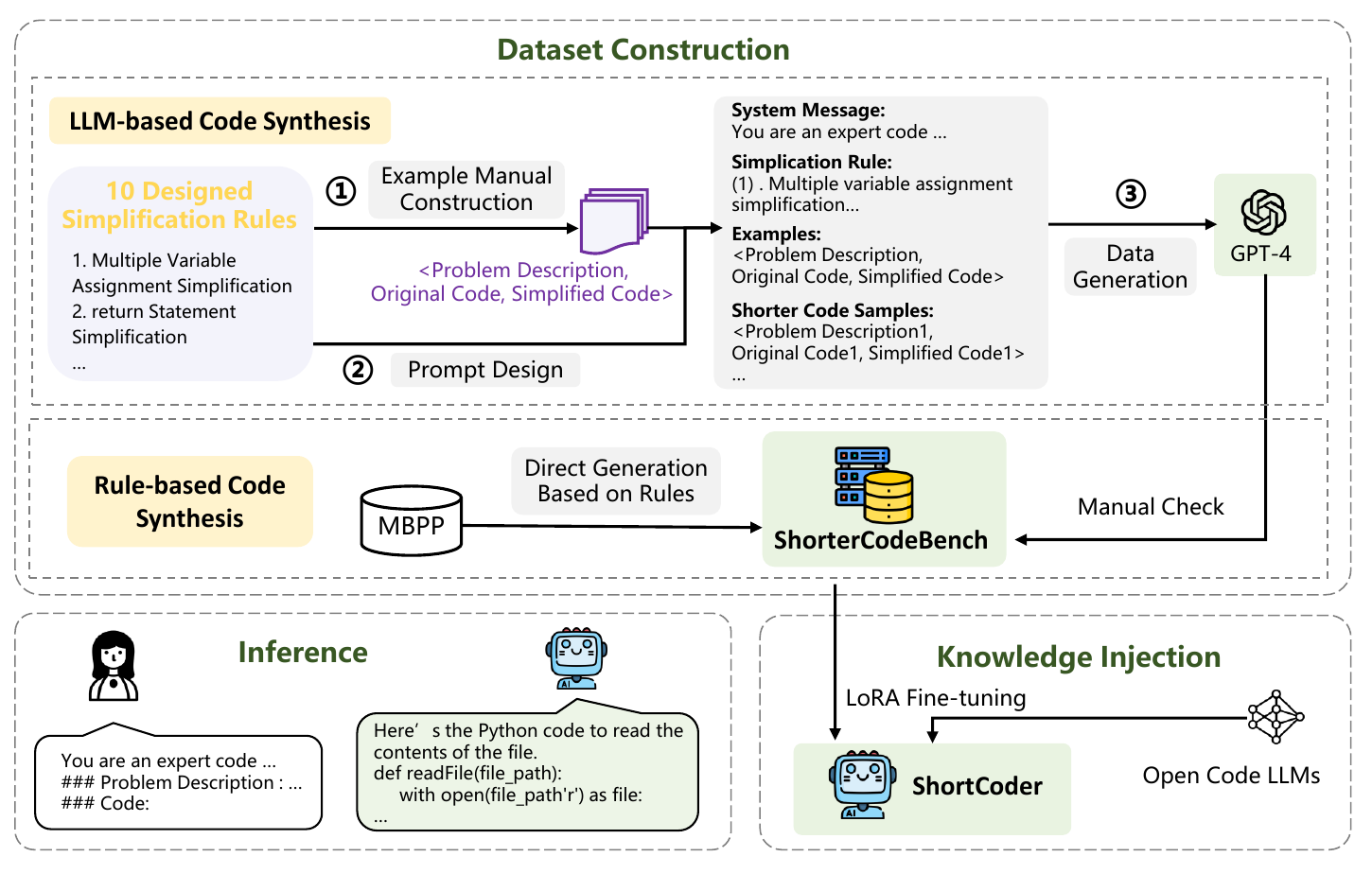}
\caption{Overview of \model.}
\label{fig:overview}
\end{figure*}

\section{Methodology}
To address the performance gap of LLMs in specialized tasks despite their strong general-domain capabilities, researchers in the natural language processing (NLP) field have proposed the concept of knowledge injection~\cite{ovadia2023fine,lauscher2020common,martino2023knowledge,cadeddu2024comparative}, which enhances task-specific performance by infusing structured or unstructured knowledge into LLMs. A mainstream implementation involves fine-tuning models with high-quality domain-specific/task-oriented data (The specific introduction of knowledge injection technology is given in Section~\ref{sec:bg_kinjection}). Inspired by these advancements, we propose a novel knowledge injection-based approach for efficient code generation, aiming to improve code generation efficiency while reducing processing latency and computational costs. Figure~\ref{fig:overview} delineates the overall framework of our technical solution. This section elaborates on the methodological components, including: (1) design principles for code simplification rules; (2) construction of knowledge-augmented datasets; (3) knowledge injection protocols, and (4) inference of \model for practical applications.

\input{rules}

\subsection{Simplification Rules Design}
\label{sec:rules}
To construct high-quality concise code corpora and enhance LLMs' code generation capability through knowledge infusion, we first designed syntax simplification rules that optimize code length while preserving semantic equivalence and readability. The rule formulation process comprises three phases:
\begin{itemize}
    \item \textbf{Manual Rule Elicitation.} We manually conduct inspection of each MBPP dataset entry by multiple authors to identify syntax-level simplification opportunities. We obtain the initial six rules through consensus-based adjudication of annotation discrepancies (Simplification Rules 1 to 6 in Table~\ref{tab:rules}).
    \item \textbf{Expert-Driven Rule Expansion.} We engage 3 experts who are proficient in Python syntax to systematically extend the rule set against Python's official documentation. We add 4 specialized rules addressing advanced language features (Simplification Rules 7 to 10 in Table~\ref{tab:rules}).
    \item \textbf{Validation-Driven Rule Finalization.} We perform cross-dataset validation using equivalent annotation protocols and confirm rule completeness through negative sampling, culminating in 10 canonical rules (As shown in Table~\ref{tab:rules}).
\end{itemize}

Finally, we obtained 10 simplification rules. The introduction of each simplification rule is as follows:

\subsubsection{Multiple Variable Assignment Simplification} 
\begin{figure}[t]
\centering
\includegraphics[width=0.95\linewidth]
{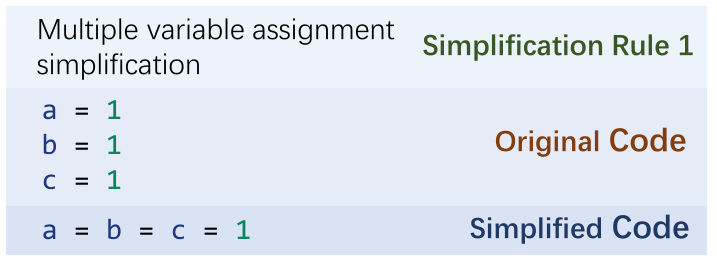}
\caption{An example illustrating the multiple variable assignment simplification rule.}
\label{fig:rule1}
\end{figure}

Simplify multiple consecutive variable assignment statements into one sentence. As shown in Figure~\ref{fig:rule1}, when multiple consecutive statements assign identical values to different variables, the rule semantically collapses them into a single assignment operation.

\subsubsection{\texttt{return} Statement Simplification} remove the parentheses in the \texttt{return} statement, enhancing code conciseness without altering program behavior. 
For example, \texttt{return ( x + y )} is simplified to \texttt{return x + y}.

\subsubsection{Assignment operation simplification} Replace verbose expressions like \texttt{x = x $\left \langle op \right \rangle$ y} with concise equivalents like x $\left \langle op \right \rangle$= y to streamline code. Note that $\left \langle op \right \rangle$ stands for assignment operators such as `` + 
 '', `` - '', `` * '', `` / '', etc.

\subsubsection{Conditional Statement Simplification} 
\begin{figure}[t]
\centering
\includegraphics[width=\linewidth]{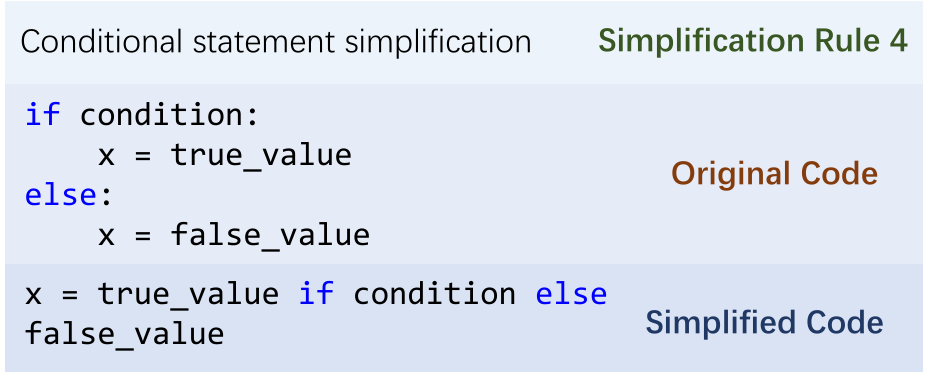}
\caption{An example illustrating the conditional statement simplification rule.}
\label{fig:rule4}
\end{figure}

focuses on simplifying single-condition branching structures by transforming verbose \texttt{if-else} statements into concise one-liner expressions. As illustrated in Figure~\ref{fig:rule4}, instead of explicitly assigning values within both the \texttt{if} and \texttt{else} blocks, the rule replaces them with a single line using a conditional (ternary) expression. This not only reduces the number of lines in the code but also enhances clarity and readability, particularly in cases where the conditional logic is simple and easily interpretable. 
Such simplification leads to more succinct and expressive code without sacrificing correctness.

\subsubsection{Multi-conditional Statement Simplification}

\begin{figure}[t]
\centering
\includegraphics[width=0.9\linewidth]{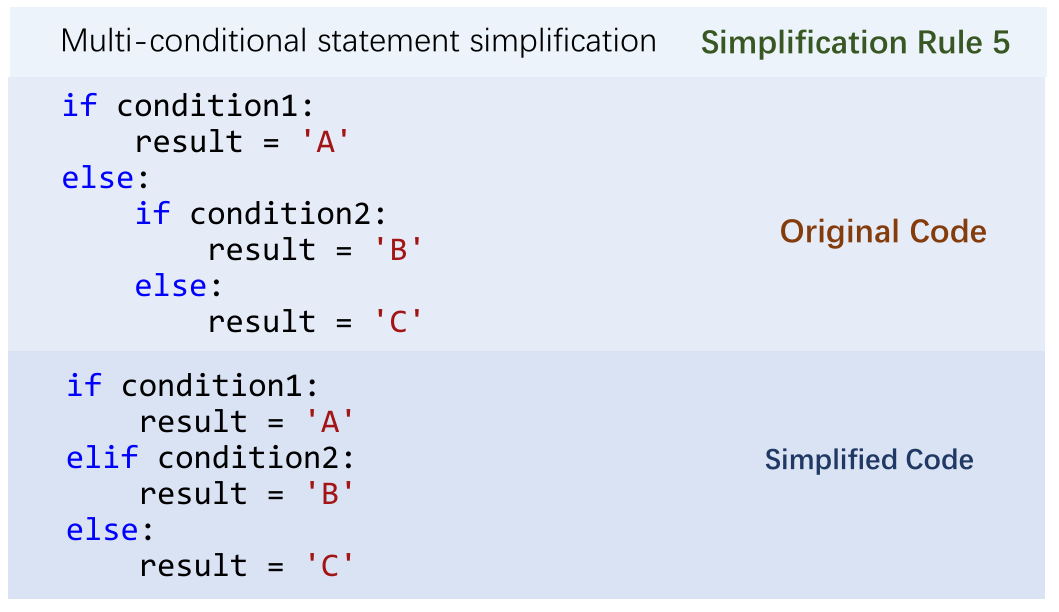}
\caption{An example illustrating multi-conditional statement simplification rule.}
\label{fig:rule5}
\end{figure}

This rule focuses on improving the readability and maintainability of the code by transforming deeply nested \texttt{if-else} statements into a more concise and linear structure using the \texttt{elif} keyword. Deeply nested conditionals can obscure logic and hinder comprehension, especially in functions with multiple decision branches. By replacing such constructs with \texttt{elif}, the logic becomes more explicit and easier to follow. Figure~\ref{fig:rule5} presents a concrete example illustrating the application of this rule.

\subsubsection{\texttt{for} Loops Simplification}

\begin{figure}[]
\centering
\includegraphics[width=\linewidth]{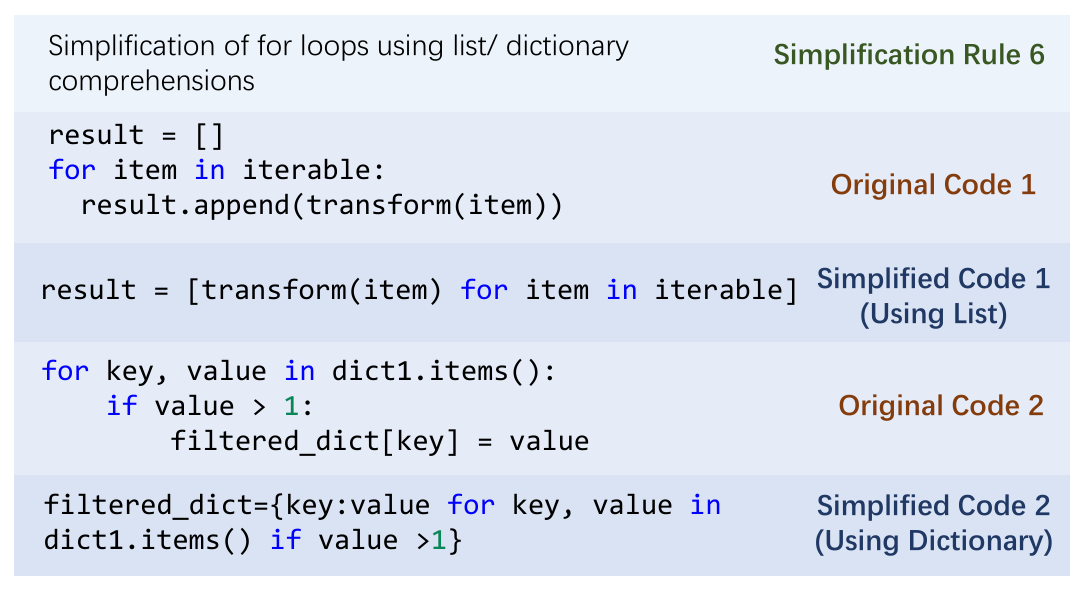}
\caption{An example illustrating \texttt{for} loops simplification rule.}
\label{fig:rule6}
\end{figure}

This rule aims to improve the code conciseness and readability by replacing traditional \texttt{for} loops with list or dictionary comprehensions where appropriate. In Python, comprehensions offer a more compact syntax for generating lists, sets, or dictionaries by embedding the loop and conditional logic within a single expression. Compared to multi-line loop constructs, comprehensions reduce boilerplate code and minimize the number of lines. Furthermore, they often lead to faster execution due to internal optimizations. Figure~\ref{fig:rule6} presents representative examples of this rule applied to both list and dictionary generation tasks, demonstrating how verbose iteration constructs can be transformed into more elegant and Pythonic expressions.

\subsubsection{Simplified Removal of Multiple Object References}

\begin{figure}[H]
\centering
\includegraphics[width=\linewidth]{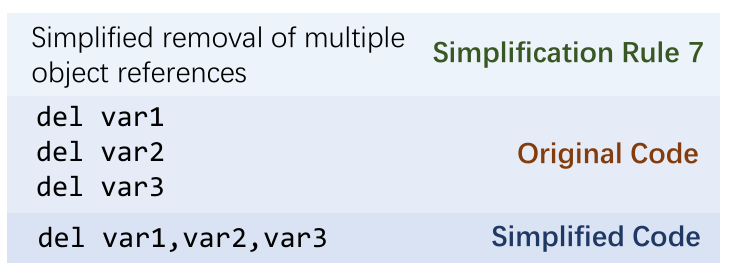}
\caption{An example illustrating simplified removal of multiple object references rule.}
\label{fig:rule7}
\end{figure}

This rule addresses the simplification of repeated \texttt{del} statements used to delete multiple object references. In Python, it is common to release references to variables by using the \texttt{del} keyword. However, writing multiple \texttt{del} statements in succession can lead to unnecessarily verbose code and reduced readability. This rule encourages the consolidation of such statements into a single line, which achieves the same functionality in a more concise and structured form. By grouping multiple deletions, the code becomes shorter and easier to understand, without compromising clarity. Figure~\ref{fig:rule7} illustrates this rule with an example where several individual \texttt{del} statements are replaced by a unified form using a comma-separated list of variables.

\subsubsection{Dictionary Mapping Simplification}

\begin{figure}[H]
\centering
\includegraphics[width=\linewidth]{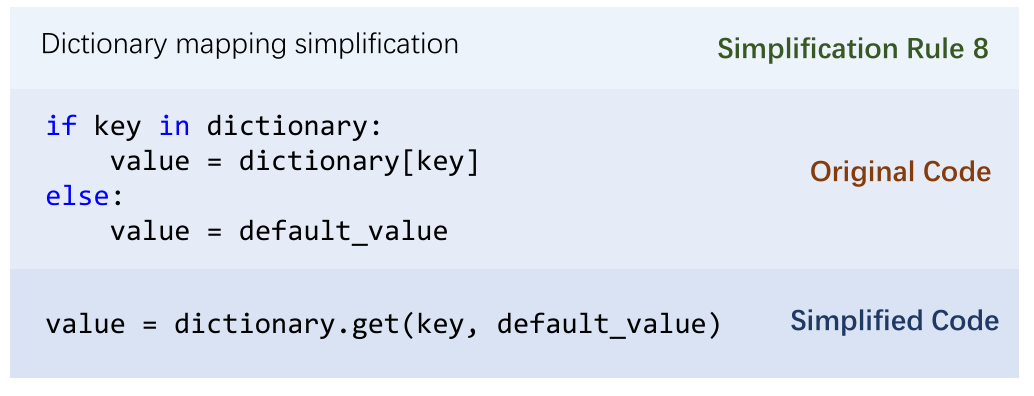}
\caption{An example illustrating dictionary mapping simplification rule.}
\label{fig:rule8}
\end{figure}

This rule simplifies key-based lookups in dictionaries by replacing verbose \texttt{if-else} structures with the more concise and idiomatic \texttt{dict.get()} method. In Python, checking whether a key exists in a dictionary before accessing its value is a common pattern, typically implemented using an explicit \texttt{if-else} block. While functional, this approach introduces unnecessary verbosity and can obscure the underlying logic.
The \texttt{get()} method provides a more readable and Pythonic alternative. Specifically, \texttt{dict.get(key, default)} returns the value associated with \texttt{key} if it exists in the dictionary; otherwise, it returns the specified \texttt{default} value. This eliminates the need for manual key existence checks using \texttt{if key in dict} and results in cleaner, more compact code.
As shown in Figure~\ref{fig:rule8}, the original implementation using a conditional check is replaced by a one-line expression with \texttt{dict.get()}, demonstrating how this simplification improves both clarity and brevity in key-based dictionary access.

\subsubsection{String Formatting Simplification}

This rule replaces verbose string concatenation with the more elegant and flexible \texttt{str.format()} method. In Python, concatenating multiple string segments using the \texttt{+} operator can quickly become cumbersome and error-prone, especially when combining strings with non-string data types that require explicit conversion. The \texttt{str.format()} method provides a structured way to embed variables and expressions within string templates, allowing for clearer and more concise formatting.


\begin{figure}[H]
\centering
\includegraphics[width=\linewidth]{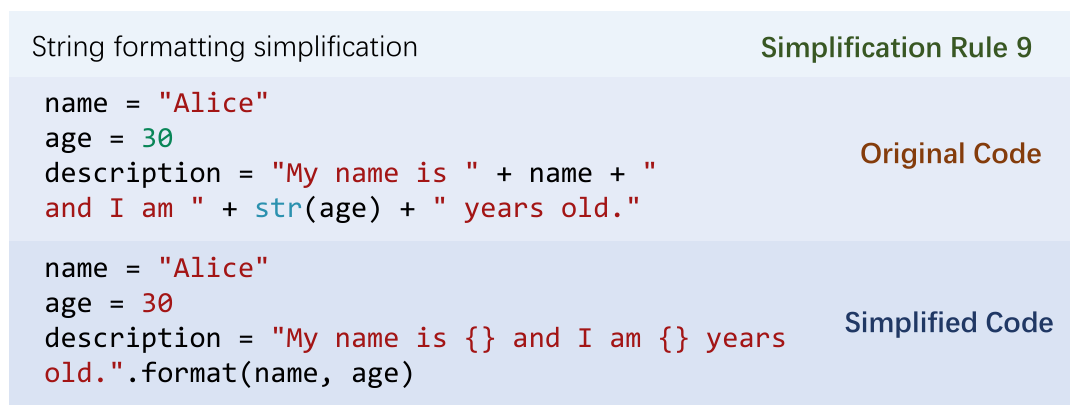}
\caption{An example illustrating string formatting simplification rule.}
\label{fig:rule9}
\end{figure}

By using \texttt{str.format()}, the code becomes easier to read and less cluttered, as the format placeholders separate the template structure from the data, facilitating both comprehension and modification. Moreover, this method supports advanced formatting options such as padding, alignment, and precision control, which are difficult to achieve cleanly with simple concatenation.

By using \texttt{str.format()}, the code becomes easier to read and less cluttered, as the format placeholders separate the template structure from the data, facilitating both comprehension and modification. Moreover, this method supports advanced formatting options such as padding, alignment, and precision control, which are difficult to achieve cleanly with simple concatenation.
Figure~\ref{fig:rule9} illustrates this simplification rule with a concrete example, demonstrating how a concatenated string expression is transformed into a more readable and maintainable formatted string using \texttt{str.format()}.

\subsubsection{File Reading and Writing Simplification}

\begin{figure}[H]
\centering
\includegraphics[width=\linewidth]{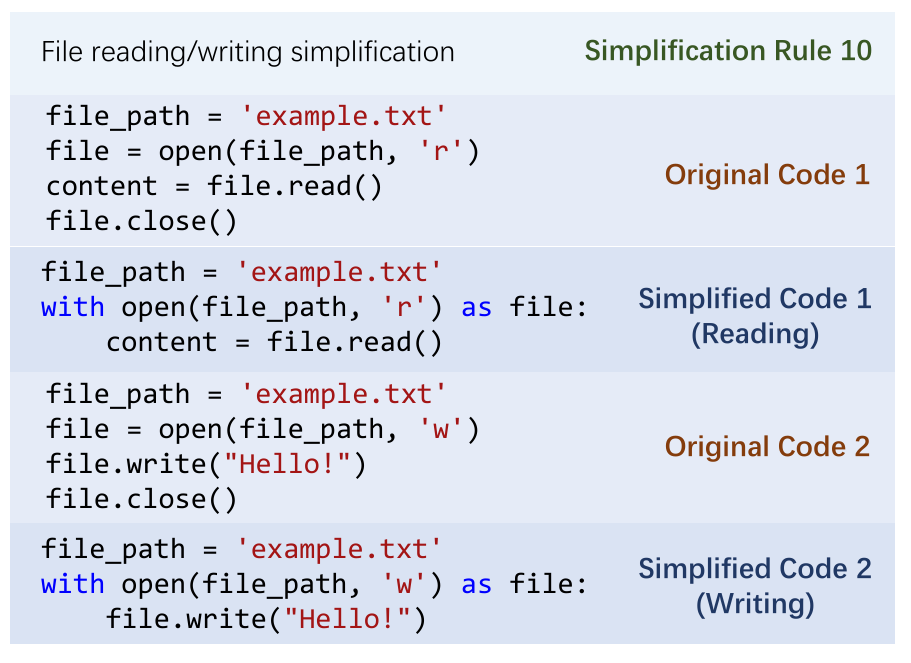}
\caption{An example illustrating file reading and writing simplification rule.}
\label{fig:rule10}
\end{figure}

This rule simplifies file I/O operations in Python by encouraging the use of the \texttt{with open()} statement instead of the traditional approach involving separate \texttt{open()} and \texttt{close()} calls. The conventional pattern requires explicitly opening a file and manually closing it after the read or write operations, which can lead to resource management issues such as unclosed file handles, particularly in the presence of exceptions or early returns.
The \texttt{with open()} statement, also known as a context manager, provides a more robust and concise alternative. It automatically handles the setup and teardown of file resources, ensuring that the file is properly closed once the block of code is exited—regardless of whether it exits normally or due to an error. This not only reduces the amount of code but also promotes safer and more reliable file handling practices.
In addition to improving safety, the \texttt{with open()} syntax enhances readability by clearly delineating the scope in which the file is used. Figure~\ref{fig:rule10} presents a representative example that contrasts the traditional verbose file operation pattern with its simplified form using the \texttt{with} statement.

\subsection{Data Construction}

\subsubsection{Rule-based Code Composition Strategy}
\begin{figure}[t]
\centering
\includegraphics[width=\linewidth]{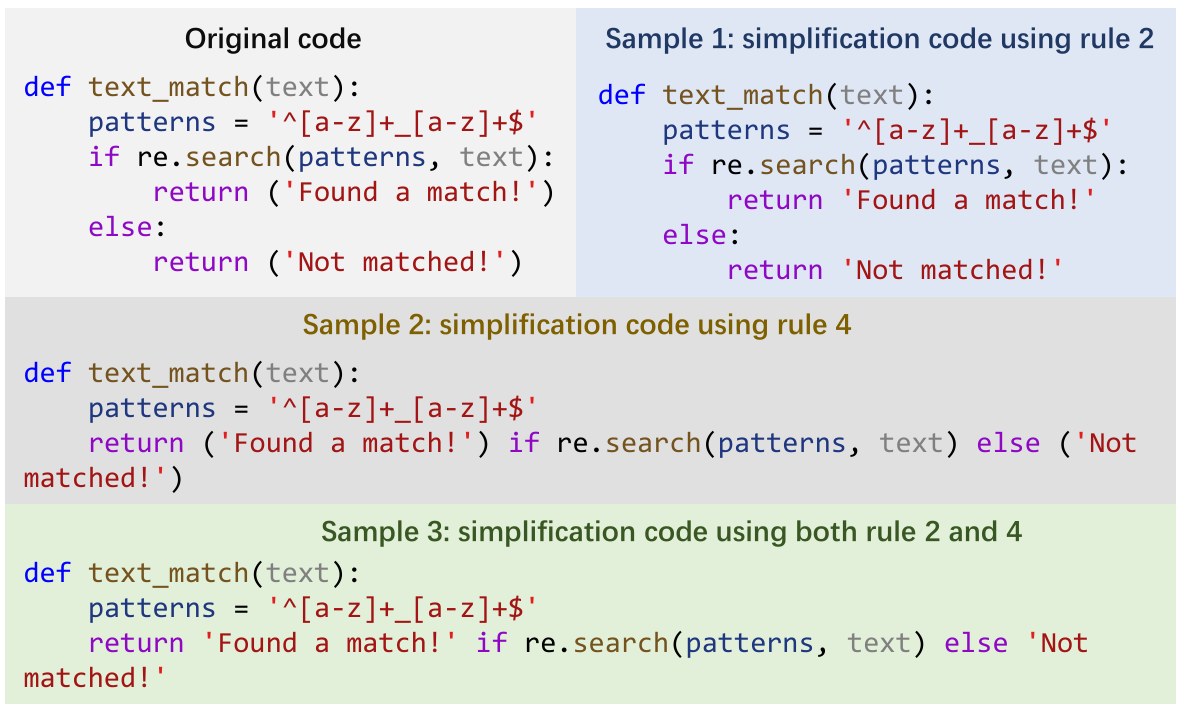}
\caption{An example applicable to multiple simplification rules.}
\label{fig:multi_rules}
\end{figure}

Based on the aforementioned simplification rules, we propose a rule-based code synthesis strategy. This strategy leverages the defined rules to transform the original code samples into their simplified counterparts. Specifically, in cases where a code snippet can be simplified by multiple rules, we introduce two construction paradigms for generating training data:
\begin{itemize}
    \item \textbf{Independent Simplification.} Each simplification rule is applied separately to the same original code, resulting in multiple simplified variants. As illustrated in Figure~\ref{fig:multi_rules}, applying Rule 2 and Rule 4 independently yields Sample 1 and Sample 2, respectively.
    \item \textbf{Joint Simplification.} Multiple simplification rules are applied simultaneously to produce a single, more concise version of the code. As shown in Figure~\ref{fig:multi_rules}, applying Rule 2 and Rule 4 together results in Sample 3.
\end{itemize}
 
Following this strategy, we construct a total of 596 pairs of $\left \langle original\_code, simplified\_code \right \rangle$ samples to support our model training and evaluation.

\subsubsection{LLM-based Code Composition Strategy}

Through manual inspection of the MBPP~\cite{austin2021program} dataset, we observe that not all predefined simplification rules are directly applicable to existing code samples. For example, with respect to Simplification Rule 10, which targets file reading and writing, the majority of programming tasks in MBPP do not involve file I/O operations. As a result, it is infeasible to construct simplified code examples using our Rule-based Synthesis Strategy.

\begin{figure}[t]
\centering
\includegraphics[width=\linewidth]{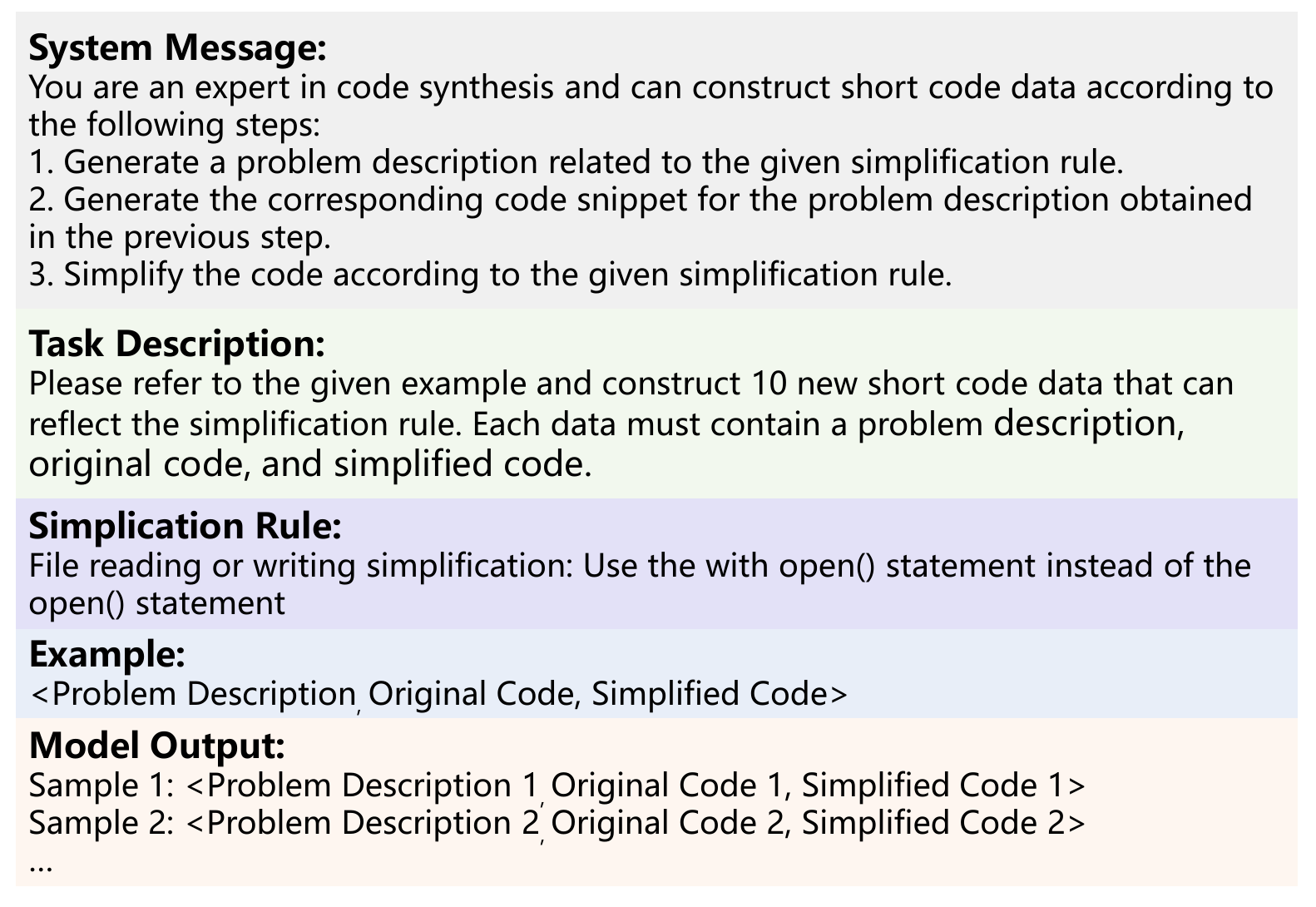}
\caption{Prompt of LLM-based code composition strategy.}
\label{fig:prompt}
\end{figure}

To address this limitation, we propose an LLM-based Code Synthesis Strategy, which leverages the capabilities of LLMs to generate simplified code tailored to specific simplification rules. The synthesis process proceeds as follows:
\begin{itemize}
    \item \textbf{Example Manual Construction.} Few-shot Learning~\cite{brown2020language} enhances the ability of language models to discern the characteristics and patterns of a task by fine-tuning it on a small number of samples, achieving effective generalization to new tasks. Inspired by this, we manually construct exemplary critiques based on the simplification rules and add them to the prompt, allowing LLMs to learn the pattern of synthesized data.
    
    \item \textbf{Prompt Design.} Following the existing work~\cite{gao2023makes, li2023large}, we construct and try various prompting patterns, ultimately adopting a prompt consisting of the task description, simplification rules, and example. Specifically, as the prompt template shown in Figure~\ref{fig:prompt}, we first assign a role and describe the pipeline of code synthesis to LLMs. Then, we provide the description of the code synthesis task and ten simplification rules. Finally, we provide an example of $\left \langle original\_code, simplified\_code \right \rangle$ pairs. Following the guide on prompt engineering provided by OpenAI\footnote{https://platform.openai.com/docs/guides/prompt-engineering/tactic-use-delimiters-to-clearly-indicate-distinct-parts-of-the-input}, we use delimiters (i.e., \{$\left \langle \ \right \rangle$, $\left \langle / \right \rangle$\}) to separate the system message, task description, simplification rules, and example in prompt.
    
    \item \textbf{Sample Generation.} Based on the designed prompts, we use them as input to synthetic code samples using LLMs. In this study, the code is synthesized using GPT-4, currently one of the most powerful LLMs available. 
\end{itemize}

\subsection{Knowledge Injection by Fine-tuning Model}
We propose a knowledge injection strategy via supervised fine-tuning to enable LLMs to generate not only functionally correct but also syntactically concise code. Specifically, we utilize the constructed dataset of $\left \langle original\_code, simplified\_code \right \rangle$ pairs derived from various code simplification rules as training data. This dataset explicitly encodes patterns of code conciseness and best practices for cleaner implementations.

To enhance the efficiency and scalability of the fine-tuning process, we adopt LoRA (Low-Rank Adaptation), a parameter-efficient fine-tuning technique~\cite{hu2022lora,xu2023parameter,han2024parameter}. LoRA injects trainable low-rank matrices into pre-trained model weights while keeping the original weights frozen. This significantly reduces the number of trainable parameters and computational overhead, making fine-tuning feasible even for LLMs on limited hardware resources.
During fine-tuning, \model learns to internalize simplification patterns and generalizes them to code generation tasks. By leveraging LoRA, we ensure that the model retains its general capabilities while effectively acquiring domain-specific knowledge related to code simplicity. As a result, the fine-tuned model \model produces high-quality, concise code with improved efficiency in both inference and generation.

\subsection{Model Inference}
The fine-tuned model is capable of performing concise code generation directly from natural language problem descriptions after fine-tuning, without requiring additional instructions or examples related to simplification rules. This contrasts with prompt engineering methods, which often rely on complex, manually designed prompts or few-shot exemplars to elicit specific coding styles or transformations.

As shown in the bottom of Figure~\ref{fig:overview}, during inference, the user provides only the target problem description as a natural language prompt. \model, having internalized knowledge of code simplification patterns during training, implicitly applies the appropriate transformation rules and produces concise, functionally correct solutions. This enables a zero-shot simplification behavior, where the model not only solves the programming task but does so in a syntactically efficient and readable manner.

\section{Experimental Setup}
We compare \model with other baselines on \dataset to evaluate the performance and efficiency of \model and design a series of ablation experiments to analyze the impact of different code synthesis strategy. In addition, we conduct human evaluation to evaluate the quality of the code generated by \model and other baselines. In this section, we will introduce the details of the experimental design, including baselines, datasets, evaluation metrics, and implementation details.

\subsection{Baselines}~\label{sec:baselines}
To evaluate the effectiveness of~\model, we select a set of representative code generation models as baselines. These models vary in architecture, scale, and pretraining strategies, providing a comprehensive view of performance across different capabilities:
\begin{itemize}
    \item \textbf{\codellama}~\cite{roziere2023code} is a state-of-the-art open-source LLM tailored for code-related tasks. It builds upon the LLaMA~\cite{touvron2023llama} architecture and is trained with code-specific data, including Python, C++, and other programming languages. The 7B variant offers a strong trade-off between performance and computational cost.
    \item \textbf{\deepseek}~\cite{guo2024deepseekcoder} is the base version of the DeepSeek-Coder model, comprising 1.3 billion parameters. Despite its relatively small size (1.3B parameters), it demonstrates competitive performance and is particularly suitable for efficiency-critical applications.
    \item \textbf{\codegen}~\cite{nijkamp2022codegen} is a family of autoregressive transformer models designed specifically for code generation, trained on large-scale code corpora and natural language descriptions. We use the multi-language version of CodeGen that supports Python, allowing us to evaluate its ability to produce concise code under standard task settings.
\end{itemize}

\subsection{Datasets}
To evaluate the effectiveness of our fine-tuned model with baseline models, we conduct experiments on the widely used HumanEval benchmark~\cite{chen2021evaluating}, a high-quality evaluation dataset designed specifically for assessing the functional correctness of code generated by LLMs.
HumanEval consists of 164 Python programming problems, each defined by a natural language prompt and a reference function signature. Crucially, each problem includes one or more test cases used to automatically validate the correctness of generated solutions.

\subsection{Evaluation Metrics}

\subsubsection{For Accuracy}
To evaluate the performance, we adopt the \textbf{pass@k} metric~\cite{chen2021evaluating}, which is widely used in previous code generation studies. The pass@k metric quantifies the probability that at least one of the top-$k$ generated code samples correctly solves a given problem according to hidden test cases.

Specially, we define the indicator function $\mathbb{I}_{i,j}$ as:

\[
\mathbb{I}_{i,j} = 
\begin{cases}
1, & \text{if the } j\text{-th sample of problem } i \text{ passes all tests} \\
0, & \text{otherwise}
\end{cases}
\]

Therefore, the pass@k metric is computed as:

  \begin{equation}
    \label{eq:passk}
   \text{pass@}k = \frac{1}{N} \sum_{i=1}^{N} \mathbb{I}\left( \sum_{j=1}^{k} \mathbb{I}_{i,j} \geq 1 \right)
    \end{equation}

where $N$ be the total number of problems in the evaluation set, and for each problem $i$, let $G_i$ be the set of $k$ generated samples. 

In our experiments, we report \textbf{pass@1}, \textbf{pass@10}, and \textbf{pass@100} to comprehensively evaluate the models’ effectiveness at different sampling budgets.

\subsubsection{For Efficiency}
To comprehensively evaluate the efficiency, we propose and employ a set of quantitative metrics that capture the resource consumption during inference. Specifically, we report the following four metrics:

\begin{itemize}
    \item \textbf{GeneratedTokens:} calculates the average number of tokens produced by the model when generating code solutions.
 
    \item \textbf{InputTokens:} calculates the average number of tokens provided to the model as input prompts for each problem.
    
    \item \textbf{TotalTokens:} is the sum of input and generated tokens, reflecting the total computational workload per problem. The specific calculation formula is as follows:
    \begin{equation}
    \label{eq:total_tokens}
    \text{TotalTokens} =\text{GeneratedTokens}+ \text{InputTokens}
    \end{equation}
    
    \item \textbf{Cost/Problem:} calculates the average cost associated with solving a single problem. This cost can be interpreted in multiple ways, such as monetary cost (e.g., API usage fees), average latency (inference time), or memory consumption. In this study, we adopt inference time as the cost metric, providing a practical estimate of the real-time efficiency of different models. The specific calculation formula is as follows:
    \begin{equation}
    \label{eq:cost_query}
    \text{Cost/Problem} = \frac{\text{Time}}{\text{Number of Problem}}
    \end{equation}
\end{itemize}

\subsection{Implementation Details}

To efficiently inject code simplification knowledge into the base model, we fine-tune it using the parameter-efficient fine-tuning technique \textbf{LoRA} (Based on \codellama). In our experiments, we set the LoRA rank to 8 and the scaling factor (LoRA alpha) to 16, which balances parameter reduction and model expressiveness. We apply LoRA to the attention projection layers, specifically targeting the \texttt{q\_proj} and \texttt{v\_proj} matrices. A dropout rate of 0.05 is used to improve generalization.

During training, we use a per-device batch size of 1 and accumulate gradients over 4 steps, resulting in an effective batch size of 4. The model is fine-tuned for 5 epochs with a learning rate of $5 \times e^{-5}$, optimized using the AdamW optimizer. To accelerate training and reduce memory consumption, we enable mixed-precision training (\texttt{fp16}).

All experiments are conducted using a single GPU with sufficient memory to support efficient fine-tuning. This setup allows for rapid adaptation of the model to our simplification task with minimal computational overhead.

\section{Evaluation Results}

In this section, we present the experimental results in detail and give an in-depth analysis to answer the following research questions (RQs):
\begin{itemize}
    \item \textbf{RQ1:} How does \model perform compared to baseline models?
    
    \item \textbf{RQ2:} How does the efficiency improvement of \model compare with other prompt-based code generation efficiency enhancement approaches?

    \item \textbf{RQ3:} How does the readability of shorter code generated by different methods compare?

\end{itemize}

\subsection{RQ1: Performance Comparison with Baseline Models.}
To evaluate the effectiveness and efficiency of our approach, we conduct experiments on the HumanEval benchmark, comparing our fine-tuned model against three baseline models: \codellama, \codegen, \deepseekBase, and \deepseekInstruct. The results are summarized in Table~\ref{tab:overall_performance}.

\begin{table}[t]
    \centering
    \setlength{\tabcolsep}{ 2.7pt}
    \caption{Performance comparison result with baseline models on HumanEval benchmark. GenTokens means the average token number of the code generated by different LLMs.}
    \label{tab:overall_performance}
    \begin{tabular}{lcccc}
    \toprule
    \textbf{Model} & \textbf{Pass@1} & \textbf{Pass@10} & \textbf{Pass@100} & \textbf{GenTokens}\\ 
    \midrule
    \codegen&0.429&0.538&0.620&260.34 \\
    \midrule
    \codellama& 0.474&0.579&0.650&197.80 \\
    \midrule
    \deepseekBase&0.324&0.457&0.690&171.35 \\
     \midrule
    \deepseekInstruct&0.791&0.884&0.920&186.69 \\
    \midrule
     \model &0.612&0.764&0.967&162.02\\
    \bottomrule
    \end{tabular}
\end{table}

As shown in Table~\ref{tab:overall_performance}, our fine-tuned model achieves a 29\%-56\% improvement in performance over the baseline, with the gap to the current SOTA models being minimal (less than 0.2). Notably, our model outperforms the current SOTA model in the pass@100 score. Additionally, our model demonstrates significant efficiency gains by generating fewer tokens than the baseline, resulting in a token reduction of 18.1\% to 37.8\%, demonstrating that the simplified code generation strategy not only preserves functional correctness but also enhances generation efficiency. These results validate the effectiveness of our fine-tuning strategy in producing concise and correct code with lower computational cost.

\begin{center}
    \begin{myboxc} \textbf{RQ1 Summary: } 
    \model significantly improves generation efficiency, achieving a 18.1\% reduction in average generated token while achieving comparable performance to baseline models on the HumanEval benchmark.
    \end{myboxc} 
\end{center}

\subsection{RQ2: Comparison with Prompt-based Efficiency Enhancement Techniques}

In RQ2, we explore and compare different strategies aimed at improving the efficiency of code generation. These strategies fall into two main categories: prompt engineering and fine-tuning. The prompt-based methods include: 
\begin{itemize}
    \item \textbf{Direct Prompting:} prompting the model to generate shorter code directly \textbf{($NoTuning\_raw$)}.
    \item \textbf{Rule-guided Prompting:} including explicit simplification rules in the prompt to guide the generation \textbf{($NoTuning\_rules$)}.
    \item \textbf{Example-Augmented Prompting:} augmenting Method 2 with in-context examples to enhance task understanding \textbf{($NoTuning\_rules\_examples$)}.
\end{itemize}  

Our method, based on rule-guided fine-tuning, belongs to the second category. We evaluate these techniques using the efficiency metrics introduced in Section III, including \textit{GeneratedTokens}, \textit{InputTokens}, \textit{TotalTokens}, and \textit{Cost/Problem}.  

As shown in Table~\ref{tab:rq2_results}, our method significantly outperforms prompt-based techniques across all efficiency indicators. In particular, our approach reduces the average cost per problem (measured by latency), demonstrating a more consistent and robust capability to generate concise code without requiring additional prompt engineering.

\begin{table*}[t]
    \centering
    \setlength{\tabcolsep}{3pt}
    \caption{Comparison results with prompt-based efficiency enhancement techniques.}
    \label{tab:rq2_results}
    \begin{tabular}{lcccc}
    \toprule
    \textbf{Method} & \textbf{\textit{GeneratedTokens}} & \textbf{\textit{InputTokens}} & \textbf{\textit{TotalTokens}} & \textbf{\textit{Cost/Problem}}\\ 
    \midrule
    $NoTuning\_raw$&214.04&151.86&365.9&3.32 \\
    \midrule
    $NoTuning\_rules$& 484.60&450.86&935.46&1.60 \\
    \midrule
    $NoTuning\_rules\_examples$&892.40&852.86&1745.26&2.46 \\
     \midrule
     \model(Fine-Tuning) &162.02&113.86&275.88&1.20\\
    \bottomrule
    \end{tabular}
\end{table*}

\begin{center}
    \begin{myboxc} \textbf{RQ2 Summary: } 
    Our rule-guided fine-tuning approach consistently achieves better efficiency than prompt-based methods. It eliminates the need for manual prompt design and achieves 25\%--64\% reductions in average problem latency, validating the advantage of knowledge injection through fine-tuning.
    \end{myboxc} 
\end{center}

\subsection{RQ3: Readability Comparison across Efficiency Enhancement Techniques.}
Inspired by previous works~\cite{hu2020deep,iyer2016summarizing,leclair2020improved,liu2019automatic,shi2021cast}, we conduct a human evaluation to assess the readability of code generated by different efficiency enhancement strategies, involving three key dimensions:
\begin{itemize}
    \item \textbf{Comprehensibility:} evaluates how easily a human can understand the intent and logic behind the code.
    \item \textbf{Clarity:} assesses the structure and presentation of the code, including how well it avoids ambiguity.
    \item \textbf{Correlation:} assesses whether the generated code is relevant to the problem. The more irrelevant information there is, the lower the score.
\end{itemize}

We first randomly choose 50 samples from the HumanEval benchmark
and their code generated by different efficiency improvement methods. In particular, we invite four volunteers with more than three years of software development experience and excellent English ability to be human evaluators. We provide each evaluator with content that includes a problem description and codes generated by different methods. All content sets are randomly shuffled for fairness in human evaluation.
Each evaluator is asked to score code generated by different methods.
The final score for each code is the average of the scores given by the four evaluators.

\begin{table}[ht]
    \centering
    \setlength{\tabcolsep}{2.0pt}
    \caption{Readability Comparison results across efficiency enhancement techniques.}
    \label{tab:readability}
    \begin{tabular}{lccc}
    \toprule
    \textbf{Method} & \textbf{Comprehensibility} & \textbf{Clarity} & \textbf{Correlation}\\ 
    \midrule
    $NoTuning\_raw$&2.66&2.70&2.72 \\
    \midrule
    $NoTuning\_rules$& 2.70&2.72&2.74 \\
    \midrule
    $NoTuning\_rules\_examples$ &2.44&2.53&2.54 \\
     \midrule
     \model (Fine-Tuning) &2.70&2.71&2.80\\
    \bottomrule
    \end{tabular}
\end{table}

Our method significantly outperformed the baselines across all three criteria. In particular, it consistently produced more concise and easier-to-understand code with more appropriate and informative comments. These findings suggest that our enhancement strategy not only improves functional correctness but also promotes better code readability, which is essential for practical usage and maintainability.

\begin{center}
    \begin{myboxc} \textbf{RQ3 Summary: } 
    Our method significantly outperformed the baselines across all three criteria, which indicates that our enhancement strategy not only improves functional correctness but also promotes better code readability.
    \end{myboxc} 
\end{center}

\section{Related Work}

\subsection{LLM-based Code Generation}

\revised{In recent years, several research efforts have explored the use of Large Language Models (LLMs) across various domains, including vulnerability detection~\cite{chen2025chatgpt,zhang2024demystifying,yang2024hyperion,zheng2025towards,cheshkov2023evaluation,tamberg2025harnessing,wang2024m2cvd,yang2024security,yusuf2024your}, commit message generation~\cite{tao2024kadel,tao2022large,shi2023sotana,tao2021evaluation,lopes2024commit,mandli2025comet,zhang2024automatic}, unit test generation~\cite{wang2025beyond,schafer2023empirical,siddiq2023exploring,chen2024chatunitest,yuan2023no}, code search~\cite{gong2024cosqa+,hu2022tackling,zheng2023survey,wang2023you,kondo2024improving,li2024procqa}, code summarization~\cite{ahmed2024automatic,wang2021cocosum,haldar2024analyzing,shi2023cocoast,lu2024yoda,duan2025hierarchical,li2024machines,su2024distilled,sun2023automatic,wang2024sparsecoder}, and code generation~\cite{huang2024karecoder,guo2024stop,wang2024rlcoder,zhong2023agieval,tao2024magis,wang2025agents,li2024repomincoder,jain2023llm,li2023skcoder,liu2024stall+,liu2023codegen4libs,ni2023lever,sun2023don,tipirneni2024structcoder,ugare2024improving,wang2023chatcoder,yuan2023evaluating,zhu2024hot}.} Code generation, a cornerstone of automated software development, involves synthesizing executable code from natural language descriptions, partial code snippets, or high-level specifications. With the proliferation of LLMs specialized for code (CodeLLMs), this field has experienced a paradigm shift. Models like CodeLlama~\cite{roziere2023code}, StarCoder~\cite{li2023starcoder}, and DeepSeekCoder~\cite{guo2024deepseekcoder} trained on massive code corpora—have demonstrated unprecedented capabilities in generating syntactically correct and semantically rich code. These models leverage self-supervised learning to capture intricate patterns in code, such as variable scoping, API dependencies, and design patterns, enabling tools like GitHub Copilot to assist developers in real-time coding workflows.

The success of CodeLLMs hinges on their ability to generalize across diverse programming tasks, from completing single lines of code to generating entire function bodies. For example, OpenAI’s Codex~\cite{finnie2022robots} excels at translating natural language prompts into Python scripts, while Meta’s CodeLlama~\cite{roziere2023code} supports multiple languages and code optimization tasks. However, despite these advancements, code generation remains challenged by the complexity of real-world software development scenarios, particularly in contexts requiring deep repository-wide understanding or domain-specific knowledge.

\subsection{Efficient Code Generation}
LLMs have significantly advanced code generation, enabling tools like GitHub Copilot to produce functionally correct code from natural language prompts, with models such as GPT-4 and CodeLlama achieving high accuracy on benchmarks like HumanEval. However, a critical limitation persists: the efficiency of code generated by LLMs often falls short of human-optimized solutions, a gap that hinders their utility in performance-sensitive scenarios. This inefficiency stems from multiple factors inherent to LLM training and inference: most code in public datasets prioritizes correctness over algorithmic efficiency, leading models to replicate common but suboptimal patterns like brute-force approaches; LLMs generate code incrementally, focusing on local syntactic coherence rather than global optimality, which can result in choices like $O(n^2)$ algorithms over more efficient alternatives; and these models lack explicit training objectives for execution time or memory usage, relying solely on correctness-driven supervision. For example, studies like EFFIBENCH~\cite{huang2024effibench} reveal that GPT-4-generated code has an average execution time 3.12 times slower than human-written canonical solutions, with extreme cases showing up to 13.89 times slower performance, highlighting the disparity in efficiency awareness. 

Challenges in achieving efficient code generation include LLMs’ limited awareness of algorithmic complexity, where they often fail to recognize optimal strategies such as dynamic programming or binary search, and suboptimal resource management, such as using inefficient data structures or introducing unnecessary computational overhead. Additionally, models struggle to adapt to varying input scales, with solutions that work for small datasets degrading significantly for larger inputs, as observed in the DPE framework’s EVALPERF benchmark~\cite{liu2024evaluating}. While instruction tuning and prompt engineering—such as adding “optimize for runtime” cues—have shown promise, improving efficiency by up to 3.2\% in GPT-4, these methods remain inconsistent. 

\section{Conclusion}

In this paper, we propose \model, a novel knowledge-augmented syntax optimization for token-efficient code generation. We design ten syntax-level simplification rules for Python, derived from AST-preserving transformations, achieving 18.1\% token reduction without functional compromise. We introduce a hybrid data synthesis pipeline integrating rule-based and LLM-based code composition strategy, producing \dataset, a corpus of 828 validated $\left \langle original\_code, simplified\_code  \right \rangle$ pairs with semantic consistency. Finally, we propose a fine-tuning strategy that injects conciseness awareness into base LLMs.
Experimental results indicate that \model achieves state-of-the-art performance in efficient code generation and demonstrates that \model is capable of generating high-quality code that is concise and highly readable.

%% file: rules.tex
\begin{table*}[t]
\scriptsize 
    \centering
    \setlength{\tabcolsep}{10pt}
    \caption{Description of the ten simplification rules.}
    \label{tab:rules}
    \begin{tabular}{cll}
        \toprule
         \textbf{No.}&\textbf{Simplification Rules}&\textbf{Description}\\
        \midrule
        1&Multiple variable assignment simplification&Combine multiple variable assignment statements into one statement.\\
        \midrule
        2&\textit{return} statement simplification&Remove the parentheses in the \textit{return} statement.\\
        \midrule
        3&Assignment operation simplification&Replace verbose expressions like x = x $\left \langle op \right \rangle$ y with concise equivalents like x $\left \langle op \right \rangle$= y.\\
        \midrule
        4&Conditional statement simplification&Simplify the single-condition statement. \\
        \midrule
        5&Multi-conditional statement simplification&Use \textit{elif} to simplify nested \textit{if-else} statements.\\
        \midrule
        6&\textit{for} loops Simplification& Simplify for loops by using list or dictionary comprehensions. \\ 
        \midrule
        7&Simplified removal of multiple object references&Combine multiple delete object reference statements into one statement.\\
        \midrule
        8&Dictionary mapping simplification&Use the \textit{dict.get()} method to replace \textit{if-else}.\\
        \midrule
        9&String formatting simplification&Use the \textit{str.format()} method instead of string concatenation.\\
        \midrule
        10&File reading and writing simplification&Use the \textit{with open()} statement instead of the traditional \textit{open()} and \textit{close()} methods.\\
        \bottomrule
    \end{tabular}
\end{table*}

%% file: main.bbl
\begin{thebibliography}{100}
\providecommand{\url}[1]{#1}
\csname url@samestyle\endcsname
\providecommand{\newblock}{\relax}
\providecommand{\bibinfo}[2]{#2}
\providecommand{\BIBentrySTDinterwordspacing}{\spaceskip=0pt\relax}
\providecommand{\BIBentryALTinterwordstretchfactor}{4}
\providecommand{\BIBentryALTinterwordspacing}{\spaceskip=\fontdimen2\font plus
\BIBentryALTinterwordstretchfactor\fontdimen3\font minus \fontdimen4\font\relax}
\providecommand{\BIBforeignlanguage}[2]{{%
\expandafter\ifx\csname l@#1\endcsname\relax
\typeout{** WARNING: IEEEtran.bst: No hyphenation pattern has been}%
\typeout{** loaded for the language `#1'. Using the pattern for}%
\typeout{** the default language instead.}%
\else
\language=\csname l@#1\endcsname
\fi
#2}}
\providecommand{\BIBdecl}{\relax}
\BIBdecl

\bibitem{he-etal-2024-cocost}
\BIBentryALTinterwordspacing
X.~He, J.~Zou, Y.~Lin, M.~Zhou, S.~Han, Z.~Yuan, and D.~Zhang, ``{C}o{C}o{ST}: Automatic complex code generation with online searching and correctness testing,'' in \emph{Proceedings of the 2024 Conference on Empirical Methods in Natural Language Processing}, Y.~Al-Onaizan, M.~Bansal, and Y.-N. Chen, Eds.\hskip 1em plus 0.5em minus 0.4em\relax Miami, Florida, USA: Association for Computational Linguistics, Nov. 2024, pp. 19\,433--19\,451. [Online]. Available: \url{https://aclanthology.org/2024.emnlp-main.1082/}
\BIBentrySTDinterwordspacing

\bibitem{2024arXiv240816498C}
L.~{Chen}, Q.~{Guo}, H.~{Jia}, Z.~{Zeng}, X.~{Wang}, Y.~{Xu}, J.~{Wu}, Y.~{Wang}, Q.~{Gao}, J.~{Wang}, W.~{Ye}, and S.~{Zhang}, ``{A Survey on Evaluating Large Language Models in Code Generation Tasks},'' \emph{arXiv e-prints}, p. arXiv:2408.16498, Aug. 2024.

\bibitem{guo2024stop}
L.~Guo, Y.~Wang, E.~Shi, W.~Zhong, H.~Zhang, J.~Chen, R.~Zhang, Y.~Ma, and Z.~Zheng, ``When to stop? towards efficient code generation in llms with excess token prevention,'' in \emph{Proceedings of the 33rd ACM SIGSOFT International Symposium on Software Testing and Analysis}, 2024, pp. 1073--1085.

\bibitem{2023arXiv231107989Z}
Z.~{Zhang}, C.~{Chen}, B.~{Liu}, C.~{Liao}, Z.~{Gong}, H.~{Yu}, J.~{Li}, and R.~{Wang}, ``{Unifying the Perspectives of NLP and Software Engineering: A Survey on Language Models for Code},'' \emph{arXiv e-prints}, p. arXiv:2311.07989, Nov. 2023.

\bibitem{2023arXiv231110372Z}
Z.~{Zheng}, K.~{Ning}, Y.~{Wang}, J.~{Zhang}, D.~{Zheng}, M.~{Ye}, and J.~{Chen}, ``{A Survey of Large Language Models for Code: Evolution, Benchmarking, and Future Trends},'' \emph{arXiv e-prints}, p. arXiv:2311.10372, Nov. 2023.

\bibitem{2023arXiv230810620H}
X.~{Hou}, Y.~{Zhao}, Y.~{Liu}, Z.~{Yang}, K.~{Wang}, L.~{Li}, X.~{Luo}, D.~{Lo}, J.~{Grundy}, and H.~{Wang}, ``{Large Language Models for Software Engineering: A Systematic Literature Review},'' \emph{arXiv e-prints}, p. arXiv:2308.10620, Aug. 2023.

\bibitem{2023arXiv230812950R}
B.~{Rozi{\`e}re}, J.~{Gehring}, F.~{Gloeckle}, S.~{Sootla}, I.~{Gat}, X.~E. {Tan}, Y.~{Adi}, J.~{Liu}, R.~{Sauvestre}, T.~{Remez}, J.~{Rapin}, A.~{Kozhevnikov}, I.~{Evtimov}, J.~{Bitton}, M.~{Bhatt}, C.~{Canton Ferrer}, A.~{Grattafiori}, W.~{Xiong}, A.~{D{\'e}fossez}, J.~{Copet}, F.~{Azhar}, H.~{Touvron}, L.~{Martin}, N.~{Usunier}, T.~{Scialom}, and G.~{Synnaeve}, ``{Code Llama: Open Foundation Models for Code},'' \emph{arXiv e-prints}, p. arXiv:2308.12950, Aug. 2023.

\bibitem{2023arXiv230506161L}
R.~{Li}, L.~{Ben Allal}, Y.~{Zi}, N.~{Muennighoff}, D.~{Kocetkov}, C.~{Mou}, M.~{Marone}, C.~{Akiki}, J.~{Li}, J.~{Chim}, Q.~{Liu}, E.~{Zheltonozhskii}, T.~Y. {Zhuo}, T.~{Wang}, O.~{Dehaene}, M.~{Davaadorj}, J.~{Lamy-Poirier}, J.~{Monteiro}, O.~{Shliazhko}, N.~{Gontier}, N.~{Meade}, A.~{Zebaze}, M.-H. {Yee}, L.~K. {Umapathi}, J.~{Zhu}, B.~{Lipkin}, M.~{Oblokulov}, Z.~{Wang}, R.~{Murthy}, J.~{Stillerman}, S.~{Sankalp Patel}, D.~{Abulkhanov}, M.~{Zocca}, M.~{Dey}, Z.~{Zhang}, N.~{Fahmy}, U.~{Bhattacharyya}, W.~{Yu}, S.~{Singh}, S.~{Luccioni}, P.~{Villegas}, M.~{Kunakov}, F.~{Zhdanov}, M.~{Romero}, T.~{Lee}, N.~{Timor}, J.~{Ding}, C.~{Schlesinger}, H.~{Schoelkopf}, J.~{Ebert}, T.~{Dao}, M.~{Mishra}, A.~{Gu}, J.~{Robinson}, C.~J. {Anderson}, B.~{Dolan-Gavitt}, D.~{Contractor}, S.~{Reddy}, D.~{Fried}, D.~{Bahdanau}, Y.~{Jernite}, C.~{Mu{\~n}oz Ferrandis}, S.~{Hughes}, T.~{Wolf}, A.~{Guha}, L.~{von Werra}, and H.~{de Vries}, ``{StarCoder: may the source be with you!}'' \emph{arXiv e-prints}, p. arXiv:2305.06161,
  May 2023.

\bibitem{jung2024discrete}
H.~Jung and K.-J. Kim, ``Discrete prompt compression with reinforcement learning,'' \emph{IEEE Access}, 2024.

\bibitem{chuang2024learning}
Y.-N. Chuang, T.~Xing, C.-Y. Chang, Z.~Liu, X.~Chen, and X.~Hu, ``Learning to compress prompt in natural language formats,'' \emph{arXiv preprint arXiv:2402.18700}, 2024.

\bibitem{huang2023fewer}
X.~Huang, L.~L. Zhang, K.-T. Cheng, F.~Yang, and M.~Yang, ``Fewer is more: Boosting llm reasoning with reinforced context pruning,'' \emph{arXiv preprint arXiv:2312.08901}, 2023.

\bibitem{2024arXiv241012388L}
Z.~{Li}, Y.~{Liu}, Y.~{Su}, and N.~{Collier}, ``{Prompt Compression for Large Language Models: A Survey},'' \emph{arXiv e-prints}, p. arXiv:2410.12388, Oct. 2024.

\bibitem{2021arXiv210313630G}
A.~{Gholami}, S.~{Kim}, Z.~{Dong}, Z.~{Yao}, M.~W. {Mahoney}, and K.~{Keutzer}, ``{A Survey of Quantization Methods for Efficient Neural Network Inference},'' \emph{arXiv e-prints}, p. arXiv:2103.13630, Mar. 2021.

\bibitem{jiang2023longllmlingua}
H.~Jiang, Q.~Wu, X.~Luo, D.~Li, C.-Y. Lin, Y.~Yang, and L.~Qiu, ``Longllmlingua: Accelerating and enhancing llms in long context scenarios via prompt compression,'' \emph{arXiv preprint arXiv:2310.06839}, 2023.

\bibitem{li2024500xcompressor}
Z.~Li, Y.~Su, and N.~Collier, ``500xcompressor: Generalized prompt compression for large language models,'' \emph{arXiv preprint arXiv:2408.03094}, 2024.

\bibitem{mu2023learning}
J.~Mu, X.~Li, and N.~Goodman, ``Learning to compress prompts with gist tokens,'' \emph{Advances in Neural Information Processing Systems}, vol.~36, pp. 19\,327--19\,352, 2023.

\bibitem{xu2023survey}
C.~Xu and J.~McAuley, ``A survey on model compression and acceleration for pretrained language models,'' in \emph{Proceedings of the AAAI Conference on Artificial Intelligence}, vol.~37, no.~9, 2023, pp. 10\,566--10\,575.

\bibitem{lang2024comprehensive}
J.~Lang, Z.~Guo, and S.~Huang, ``A comprehensive study on quantization techniques for large language models,'' in \emph{2024 4th International Conference on Artificial Intelligence, Robotics, and Communication (ICAIRC)}.\hskip 1em plus 0.5em minus 0.4em\relax IEEE, 2024, pp. 224--231.

\bibitem{wei2023greener}
X.~Wei, S.~Gonugondla, W.~Ahmad, S.~Wang, B.~Ray, H.~Qian, X.~Li, V.~Kumar, Z.~Wang, Y.~Tian \emph{et~al.}, ``Greener yet powerful: Taming large code generation models with quantization,'' \emph{arXiv preprint arXiv:2303.05378}, 2023.

\bibitem{frantar2022gptq}
E.~Frantar, S.~Ashkboos, T.~Hoefler, and D.~Alistarh, ``Gptq: Accurate post-training quantization for generative pre-trained transformers,'' \emph{arXiv preprint arXiv:2210.17323}, 2022.

\bibitem{liu2024vptq}
Y.~Liu, J.~Wen, Y.~Wang, S.~Ye, L.~L. Zhang, T.~Cao, C.~Li, and M.~Yang, ``Vptq: Extreme low-bit vector post-training quantization for large language models,'' \emph{arXiv preprint arXiv:2409.17066}, 2024.

\bibitem{2024arXiv240416333S}
Z.~{Sun}, X.~{Du}, Z.~{Yang}, L.~{Li}, and D.~{Lo}, ``{AI Coders Are Among Us: Rethinking Programming Language Grammar Towards Efficient Code Generation},'' \emph{arXiv e-prints}, p. arXiv:2404.16333, Apr. 2024.

\bibitem{wang2023codet5+}
Y.~Wang, H.~Le, A.~D. Gotmare, N.~D. Bui, J.~Li, and S.~C. Hoi, ``Codet5+: Open code large language models for code understanding and generation,'' \emph{arXiv preprint arXiv:2305.07922}, 2023.

\bibitem{murali2023codecompose}
V.~Murali, C.~Maddila, I.~Ahmad, M.~Bolin, D.~Cheng, N.~Ghorbani, R.~Fernandez, and N.~Nagappan, ``Codecompose: A large-scale industrial deployment of ai-assisted code authoring,'' \emph{arXiv preprint arXiv:2305.12050}, 2023.

\bibitem{fu2023revisiting}
P.~Fu, Y.~Zhang, H.~Wang, W.~Qiu, and J.~Zhao, ``Revisiting the knowledge injection frameworks,'' \emph{arXiv preprint arXiv:2311.01150}, 2023.

\bibitem{zan2023private}
D.~Zan, B.~Chen, Y.~Gong, J.~Cao, F.~Zhang, B.~Wu, B.~Guan, Y.~Yin, and Y.~Wang, ``Private-library-oriented code generation with large language models,'' \emph{arXiv preprint arXiv:2307.15370}, 2023.

\bibitem{fried2022incoder}
D.~Fried, A.~Aghajanyan, J.~Lin, S.~Wang, E.~Wallace, F.~Shi, R.~Zhong, W.-t. Yih, L.~Zettlemoyer, and M.~Lewis, ``Incoder: A generative model for code infilling and synthesis,'' \emph{arXiv preprint arXiv:2204.05999}, 2022.

\bibitem{ovadia2023fine}
O.~Ovadia, M.~Brief, M.~Mishaeli, and O.~Elisha, ``Fine-tuning or retrieval? comparing knowledge injection in llms,'' \emph{arXiv preprint arXiv:2312.05934}, 2023.

\bibitem{lauscher2020common}
A.~Lauscher, O.~Majewska, L.~F. Ribeiro, I.~Gurevych, N.~Rozanov, and G.~Glava{\v{s}}, ``Common sense or world knowledge? investigating adapter-based knowledge injection into pretrained transformers,'' \emph{arXiv preprint arXiv:2005.11787}, 2020.

\bibitem{martino2023knowledge}
A.~Martino, M.~Iannelli, and C.~Truong, ``Knowledge injection to counter large language model (llm) hallucination,'' in \emph{European Semantic Web Conference}.\hskip 1em plus 0.5em minus 0.4em\relax Springer, 2023, pp. 182--185.

\bibitem{cadeddu2024comparative}
A.~Cadeddu, A.~Chessa, V.~De~Leo, G.~Fenu, E.~Motta, F.~Osborne, D.~R. Recupero, A.~Salatino, and L.~Secchi, ``A comparative analysis of knowledge injection strategies for large language models in the scholarly domain,'' \emph{Engineering Applications of Artificial Intelligence}, vol. 133, p. 108166, 2024.

\bibitem{austin2021program}
J.~Austin, A.~Odena, M.~Nye, M.~Bosma, H.~Michalewski, D.~Dohan, E.~Jiang, C.~Cai, M.~Terry, Q.~Le \emph{et~al.}, ``Program synthesis with large language models,'' \emph{arXiv preprint arXiv:2108.07732}, 2021.

\bibitem{brown2020language}
T.~Brown, B.~Mann, N.~Ryder, M.~Subbiah, J.~D. Kaplan, P.~Dhariwal, A.~Neelakantan, P.~Shyam, G.~Sastry, A.~Askell \emph{et~al.}, ``Language models are few-shot learners,'' \emph{Advances in neural information processing systems}, vol.~33, pp. 1877--1901, 2020.

\bibitem{gao2023makes}
S.~Gao, X.-C. Wen, C.~Gao, W.~Wang, H.~Zhang, and M.~R. Lyu, ``What makes good in-context demonstrations for code intelligence tasks with llms?'' in \emph{2023 38th IEEE/ACM International Conference on Automated Software Engineering (ASE)}.\hskip 1em plus 0.5em minus 0.4em\relax IEEE, 2023, pp. 761--773.

\bibitem{li2023large}
J.~Li, G.~Li, C.~Tao, H.~Zhang, F.~Liu, and Z.~Jin, ``Large language model-aware in-context learning for code generation,'' \emph{arXiv preprint arXiv:2310.09748}, 2023.

\bibitem{hu2022lora}
E.~J. Hu, Y.~Shen, P.~Wallis, Z.~Allen-Zhu, Y.~Li, S.~Wang, L.~Wang, W.~Chen \emph{et~al.}, ``Lora: Low-rank adaptation of large language models.'' \emph{ICLR}, vol.~1, no.~2, p.~3, 2022.

\bibitem{xu2023parameter}
L.~Xu, H.~Xie, S.-Z.~J. Qin, X.~Tao, and F.~L. Wang, ``Parameter-efficient fine-tuning methods for pretrained language models: A critical review and assessment,'' \emph{arXiv preprint arXiv:2312.12148}, 2023.

\bibitem{han2024parameter}
Z.~Han, C.~Gao, J.~Liu, J.~Zhang, and S.~Q. Zhang, ``Parameter-efficient fine-tuning for large models: A comprehensive survey,'' \emph{arXiv preprint arXiv:2403.14608}, 2024.

\bibitem{roziere2023code}
B.~Roziere, J.~Gehring, F.~Gloeckle, S.~Sootla, I.~Gat, X.~E. Tan, Y.~Adi, J.~Liu, R.~Sauvestre, T.~Remez \emph{et~al.}, ``Code llama: Open foundation models for code,'' \emph{arXiv preprint arXiv:2308.12950}, 2023.

\bibitem{touvron2023llama}
H.~Touvron, T.~Lavril, G.~Izacard, X.~Martinet, M.-A. Lachaux, T.~Lacroix, B.~Rozi{\`e}re, N.~Goyal, E.~Hambro, F.~Azhar \emph{et~al.}, ``Llama: Open and efficient foundation language models,'' \emph{arXiv preprint arXiv:2302.13971}, 2023.

\bibitem{guo2024deepseekcoder}
D.~Guo, Q.~Zhu, D.~Yang, Z.~Xie, K.~Dong, W.~Zhang, G.~Chen, X.~Bi, Y.~Wu, Y.~Li \emph{et~al.}, ``Deepseek-coder: When the large language model meets programming--the rise of code intelligence,'' \emph{arXiv preprint arXiv:2401.14196}, 2024.

\bibitem{nijkamp2022codegen}
E.~Nijkamp, B.~Pang, H.~Hayashi, L.~Tu, H.~Wang, Y.~Zhou, S.~Savarese, and C.~Xiong, ``Codegen: An open large language model for code with multi-turn program synthesis,'' \emph{arXiv preprint arXiv:2203.13474}, 2022.

\bibitem{chen2021evaluating}
M.~Chen, J.~Tworek, H.~Jun, Q.~Yuan, H.~P. d.~O. Pinto, J.~Kaplan, H.~Edwards, Y.~Burda, N.~Joseph, G.~Brockman \emph{et~al.}, ``Evaluating large language models trained on code,'' \emph{arXiv preprint arXiv:2107.03374}, 2021.

\bibitem{hu2020deep}
X.~Hu, G.~Li, X.~Xia, D.~Lo, and Z.~Jin, ``Deep code comment generation with hybrid lexical and syntactical information,'' \emph{Empirical Software Engineering}, vol.~25, no.~3, pp. 2179--2217, 2020.

\bibitem{iyer2016summarizing}
S.~Iyer, I.~Konstas, A.~Cheung, and L.~Zettlemoyer, ``Summarizing source code using a neural attention model,'' in \emph{54th Annual Meeting of the Association for Computational Linguistics 2016}.\hskip 1em plus 0.5em minus 0.4em\relax Association for Computational Linguistics, 2016, pp. 2073--2083.

\bibitem{leclair2020improved}
A.~LeClair, S.~Haque, L.~Wu, and C.~McMillan, ``Improved code summarization via a graph neural network,'' in \emph{Proceedings of the 28th international conference on program comprehension}, 2020, pp. 184--195.

\bibitem{liu2019automatic}
T.~Liu, W.~Ding, Z.~Wang, J.~Tang, G.~Y. Huang, and Z.~Liu, ``Automatic short answer grading via multiway attention networks,'' in \emph{Artificial Intelligence in Education: 20th International Conference, AIED 2019, Chicago, IL, USA, June 25-29, 2019, Proceedings, Part II 20}.\hskip 1em plus 0.5em minus 0.4em\relax Springer, 2019, pp. 169--173.

\bibitem{shi2021cast}
E.~Shi, Y.~Wang, L.~Du, H.~Zhang, S.~Han, D.~Zhang, and H.~Sun, ``Cast: Enhancing code summarization with hierarchical splitting and reconstruction of abstract syntax trees,'' \emph{arXiv preprint arXiv:2108.12987}, 2021.

\bibitem{chen2025chatgpt}
C.~Chen, J.~Su, J.~Chen, Y.~Wang, T.~Bi, J.~Yu, Y.~Wang, X.~Lin, T.~Chen, and Z.~Zheng, ``When chatgpt meets smart contract vulnerability detection: How far are we?'' \emph{ACM Transactions on Software Engineering and Methodology}, vol.~34, no.~4, pp. 1--30, 2025.

\bibitem{zhang2024demystifying}
J.~Zhang, Y.~Shen, J.~Chen, J.~Su, Y.~Wang, T.~Chen, J.~Gao, and Z.~Chen, ``Demystifying and detecting cryptographic defects in ethereum smart contracts,'' \emph{arXiv preprint arXiv:2408.04939}, 2024.

\bibitem{yang2024hyperion}
S.~Yang, X.~Lin, J.~Chen, Q.~Zhong, L.~Xiao, R.~Huang, Y.~Wang, and Z.~Zheng, ``Hyperion: Unveiling dapp inconsistencies using llm and dataflow-guided symbolic execution,'' \emph{arXiv preprint arXiv:2408.06037}, 2024.

\bibitem{zheng2025towards}
Z.~Zheng, K.~Ning, Q.~Zhong, J.~Chen, W.~Chen, L.~Guo, W.~Wang, and Y.~Wang, ``Towards an understanding of large language models in software engineering tasks,'' \emph{Empirical Software Engineering}, vol.~30, no.~2, p.~50, 2025.

\bibitem{cheshkov2023evaluation}
A.~Cheshkov, P.~Zadorozhny, and R.~Levichev, ``Evaluation of chatgpt model for vulnerability detection,'' \emph{arXiv preprint arXiv:2304.07232}, 2023.

\bibitem{tamberg2025harnessing}
K.~Tamberg and H.~Bahsi, ``Harnessing large language models for software vulnerability detection: A comprehensive benchmarking study,'' \emph{IEEE Access}, 2025.

\bibitem{wang2024m2cvd}
Z.~Wang, G.~Li, J.~Li, Y.~Xiong, M.~Yan, and Z.~Jin, ``M2cvd: Enhancing vulnerability semantic through multi-model collaboration for code vulnerability detection,'' \emph{arXiv preprint arXiv:2406.05940}, 2024.

\bibitem{yang2024security}
A.~Z. Yang, H.~Tian, H.~Ye, R.~Martins, and C.~L. Goues, ``Security vulnerability detection with multitask self-instructed fine-tuning of large language models,'' \emph{arXiv preprint arXiv:2406.05892}, 2024.

\bibitem{yusuf2024your}
I.~N.~B. Yusuf and L.~Jiang, ``Your instructions are not always helpful: Assessing the efficacy of instruction fine-tuning for software vulnerability detection,'' \emph{arXiv preprint arXiv:2401.07466}, 2024.

\bibitem{tao2024kadel}
W.~Tao, Y.~Zhou, Y.~Wang, H.~Zhang, H.~Wang, and W.~Zhang, ``Kadel: Knowledge-aware denoising learning for commit message generation,'' \emph{ACM Transactions on Software Engineering and Methodology}, vol.~33, no.~5, pp. 1--32, 2024.

\bibitem{tao2022large}
W.~Tao, Y.~Wang, E.~Shi, L.~Du, S.~Han, H.~Zhang, D.~Zhang, and W.~Zhang, ``A large-scale empirical study of commit message generation: models, datasets and evaluation,'' \emph{Empirical Software Engineering}, vol.~27, no.~7, p. 198, 2022.

\bibitem{shi2023sotana}
E.~Shi, F.~Zhang, Y.~Wang, B.~Chen, L.~Du, H.~Zhang, S.~Han, D.~Zhang, and H.~Sun, ``Sotana: The open-source software development assistant,'' \emph{arXiv preprint arXiv:2308.13416}, 2023.

\bibitem{tao2021evaluation}
W.~Tao, Y.~Wang, E.~Shi, L.~Du, S.~Han, H.~Zhang, D.~Zhang, and W.~Zhang, ``On the evaluation of commit message generation models: An experimental study,'' in \emph{2021 IEEE International Conference on Software Maintenance and Evolution (ICSME)}.\hskip 1em plus 0.5em minus 0.4em\relax IEEE, 2021, pp. 126--136.

\bibitem{lopes2024commit}
C.~V. Lopes, V.~I. Klotzman, I.~Ma, and I.~Ahmed, ``Commit messages in the age of large language models,'' \emph{arXiv preprint arXiv:2401.17622}, 2024.

\bibitem{mandli2025comet}
A.~R. Mandli, S.~Rajput, and T.~Sharma, ``Comet: Generating commit messages using delta graph context representation,'' \emph{Journal of Systems and Software}, vol. 222, p. 112307, 2025.

\bibitem{zhang2024automatic}
Y.~Zhang, Z.~Qiu, K.-J. Stol, W.~Zhu, J.~Zhu, Y.~Tian, and H.~Liu, ``Automatic commit message generation: A critical review and directions for future work,'' \emph{IEEE Transactions on Software Engineering}, vol.~50, no.~4, pp. 816--835, 2024.

\bibitem{wang2025beyond}
Y.~Wang, T.~Jiang, M.~Liu, J.~Chen, M.~Mao, X.~Liu, Y.~Ma, and Z.~Zheng, ``Beyond functional correctness: Investigating coding style inconsistencies in large language models,'' \emph{Proceedings of the ACM on Software Engineering}, vol.~2, no. FSE, pp. 690--712, 2025.

\bibitem{schafer2023empirical}
M.~Sch{\"a}fer, S.~Nadi, A.~Eghbali, and F.~Tip, ``An empirical evaluation of using large language models for automated unit test generation,'' \emph{IEEE Transactions on Software Engineering}, vol.~50, no.~1, pp. 85--105, 2023.

\bibitem{siddiq2023exploring}
M.~L. Siddiq, J.~C. Santos, R.~H. Tanvir, N.~Ulfat, F.~Al~Rifat, and V.~C. Lopes, ``Exploring the effectiveness of large language models in generating unit tests,'' \emph{CoRR}, 2023.

\bibitem{chen2024chatunitest}
Y.~Chen, Z.~Hu, C.~Zhi, J.~Han, S.~Deng, and J.~Yin, ``Chatunitest: A framework for llm-based test generation,'' in \emph{Companion Proceedings of the 32nd ACM International Conference on the Foundations of Software Engineering}, 2024, pp. 572--576.

\bibitem{yuan2023no}
Z.~Yuan, Y.~Lou, M.~Liu, S.~Ding, K.~Wang, Y.~Chen, and X.~Peng, ``No more manual tests? evaluating and improving chatgpt for unit test generation,'' \emph{arXiv preprint arXiv:2305.04207}, 2023.

\bibitem{gong2024cosqa+}
J.~Gong, Y.~Wu, L.~Liang, Z.~Zheng, and Y.~Wang, ``Cosqa+: Enhancing code search dataset with matching code,'' \emph{arXiv e-prints}, pp. arXiv--2406, 2024.

\bibitem{hu2022tackling}
F.~Hu, Y.~Wang, L.~Du, H.~Zhang, S.~Han, D.~Zhang, and X.~Li, ``Tackling long code search with splitting, encoding, and aggregating,'' \emph{arXiv preprint arXiv:2208.11271}, 2022.

\bibitem{zheng2023survey}
Z.~Zheng, K.~Ning, Y.~Wang, J.~Zhang, D.~Zheng, M.~Ye, and J.~Chen, ``A survey of large language models for code: Evolution, benchmarking, and future trends,'' \emph{arXiv preprint arXiv:2311.10372}, 2023.

\bibitem{wang2023you}
Y.~Wang, L.~Guo, E.~Shi, W.~Chen, J.~Chen, W.~Zhong, M.~Wang, H.~Li, H.~Zhang, Z.~Lyu \emph{et~al.}, ``You augment me: Exploring chatgpt-based data augmentation for semantic code search,'' in \emph{2023 IEEE International Conference on Software Maintenance and Evolution (ICSME)}.\hskip 1em plus 0.5em minus 0.4em\relax IEEE, 2023, pp. 14--25.

\bibitem{kondo2024improving}
M.~Kondo, D.~Kawahara, and T.~Kurabayashi, ``Improving repository-level code search with text conversion,'' in \emph{Proceedings of the 2024 Conference of the North American Chapter of the Association for Computational Linguistics: Human Language Technologies (Volume 4: Student Research Workshop)}, 2024, pp. 130--137.

\bibitem{li2024procqa}
Z.~Li, J.~Zhang, C.~Yin, Y.~Ouyang, and W.~Rong, ``Procqa: a large-scale community-based programming question answering dataset for code search,'' \emph{arXiv preprint arXiv:2403.16702}, 2024.

\bibitem{ahmed2024automatic}
T.~Ahmed, K.~S. Pai, P.~Devanbu, and E.~T. Barr, ``Automatic semantic augmentation of language model prompts (for code summarization),'' in \emph{2024 IEEE/ACM 46th International Conference on Software Engineering (ICSE)}.\hskip 1em plus 0.5em minus 0.4em\relax IEEE Computer Society, 2024, pp. 1004--1004.

\bibitem{wang2021cocosum}
Y.~Wang, E.~Shi, L.~Du, X.~Yang, Y.~Hu, S.~Han, H.~Zhang, and D.~Zhang, ``Cocosum: Contextual code summarization with multi-relational graph neural network,'' \emph{arXiv preprint arXiv:2107.01933}, 2021.

\bibitem{haldar2024analyzing}
R.~Haldar and J.~Hockenmaier, ``Analyzing the performance of large language models on code summarization,'' \emph{arXiv preprint arXiv:2404.08018}, 2024.

\bibitem{shi2023cocoast}
E.~Shi, Y.~Wang, L.~Du, H.~Zhang, S.~Han, D.~Zhang, and H.~Sun, ``Cocoast: representing source code via hierarchical splitting and reconstruction of abstract syntax trees,'' \emph{Empirical Software Engineering}, vol.~28, no.~6, p. 135, 2023.

\bibitem{lu2024yoda}
J.~Lu, W.~Zhong, Y.~Wang, Z.~Guo, Q.~Zhu, W.~Huang, Y.~Wang, F.~Mi, B.~Wang, Y.~Wang \emph{et~al.}, ``Yoda: Teacher-student progressive learning for language models,'' \emph{arXiv preprint arXiv:2401.15670}, 2024.

\bibitem{duan2025hierarchical}
G.~Duan, M.~Liu, Y.~Wang, C.~Wang, X.~Peng, and Z.~Zheng, ``A hierarchical and evolvable benchmark for fine-grained code instruction following with multi-turn feedback,'' \emph{arXiv preprint arXiv:2507.00699}, 2025.

\bibitem{li2024machines}
J.~Li, Y.~Zhang, Z.~Karas, C.~McMillan, K.~Leach, and Y.~Huang, ``Do machines and humans focus on similar code? exploring explainability of large language models in code summarization,'' in \emph{Proceedings of the 32nd IEEE/ACM International Conference on Program Comprehension}, 2024, pp. 47--51.

\bibitem{su2024distilled}
C.-Y. Su and C.~McMillan, ``Distilled gpt for source code summarization,'' \emph{Automated Software Engineering}, vol.~31, no.~1, p.~22, 2024.

\bibitem{sun2023automatic}
W.~Sun, C.~Fang, Y.~You, Y.~Miao, Y.~Liu, Y.~Li, G.~Deng, S.~Huang, Y.~Chen, Q.~Zhang \emph{et~al.}, ``Automatic code summarization via chatgpt: How far are we?'' \emph{arXiv preprint arXiv:2305.12865}, 2023.

\bibitem{wang2024sparsecoder}
Y.~Wang, Y.~Huang, D.~Guo, H.~Zhang, and Z.~Zheng, ``Sparsecoder: Identifier-aware sparse transformer for file-level code summarization,'' in \emph{2024 IEEE International Conference on Software Analysis, Evolution and Reengineering (SANER)}.\hskip 1em plus 0.5em minus 0.4em\relax IEEE, 2024, pp. 614--625.

\bibitem{huang2024karecoder}
T.~Huang, Z.~Sun, Z.~Jin, G.~Li, and C.~Lyu, ``Karecoder: A new knowledge-enriched code generation system,'' in \emph{Proceedings of the 2024 IEEE/ACM 46th International Conference on Software Engineering: Companion Proceedings}, 2024, pp. 270--271.

\bibitem{wang2024rlcoder}
Y.~Wang, Y.~Wang, D.~Guo, J.~Chen, R.~Zhang, Y.~Ma, and Z.~Zheng, ``Rlcoder: Reinforcement learning for repository-level code completion,'' \emph{arXiv preprint arXiv:2407.19487}, 2024.

\bibitem{zhong2023agieval}
W.~Zhong, R.~Cui, Y.~Guo, Y.~Liang, S.~Lu, Y.~Wang, A.~Saied, W.~Chen, and N.~Duan, ``Agieval: A human-centric benchmark for evaluating foundation models,'' \emph{arXiv preprint arXiv:2304.06364}, 2023.

\bibitem{tao2024magis}
W.~Tao, Y.~Zhou, Y.~Wang, W.~Zhang, H.~Zhang, and Y.~Cheng, ``Magis: Llm-based multi-agent framework for github issue resolution,'' \emph{Advances in Neural Information Processing Systems}, vol.~37, pp. 51\,963--51\,993, 2024.

\bibitem{wang2025agents}
Y.~Wang, W.~Zhong, Y.~Huang, E.~Shi, M.~Yang, J.~Chen, H.~Li, Y.~Ma, Q.~Wang, and Z.~Zheng, ``Agents in software engineering: Survey, landscape, and vision,'' \emph{Automated Software Engineering}, vol.~32, no.~2, pp. 1--36, 2025.

\bibitem{li2024repomincoder}
Y.~Li, E.~Shi, D.~Zheng, K.~Duan, J.~Chen, and Y.~Wang, ``Repomincoder: Improving repository-level code generation based on information loss screening,'' in \emph{Proceedings of the 15th Asia-Pacific Symposium on Internetware}, 2024, pp. 229--238.

\bibitem{jain2023llm}
N.~Jain, T.~Zhang, W.-L. Chiang, J.~E. Gonzalez, K.~Sen, and I.~Stoica, ``Llm-assisted code cleaning for training accurate code generators,'' \emph{arXiv preprint arXiv:2311.14904}, 2023.

\bibitem{li2023skcoder}
J.~Li, Y.~Li, G.~Li, Z.~Jin, Y.~Hao, and X.~Hu, ``Skcoder: A sketch-based approach for automatic code generation,'' in \emph{2023 IEEE/ACM 45th International Conference on Software Engineering (ICSE)}.\hskip 1em plus 0.5em minus 0.4em\relax IEEE, 2023, pp. 2124--2135.

\bibitem{liu2024stall+}
J.~Liu, Y.~Chen, M.~Liu, X.~Peng, and Y.~Lou, ``Stall+: Boosting llm-based repository-level code completion with static analysis,'' \emph{arXiv preprint arXiv:2406.10018}, 2024.

\bibitem{liu2023codegen4libs}
M.~Liu, T.~Yang, Y.~Lou, X.~Du, Y.~Wang, and X.~Peng, ``Codegen4libs: A two-stage approach for library-oriented code generation,'' in \emph{2023 38th IEEE/ACM International Conference on Automated Software Engineering (ASE)}.\hskip 1em plus 0.5em minus 0.4em\relax IEEE, 2023, pp. 434--445.

\bibitem{ni2023lever}
A.~Ni, S.~Iyer, D.~Radev, V.~Stoyanov, W.-t. Yih, S.~Wang, and X.~V. Lin, ``Lever: Learning to verify language-to-code generation with execution,'' in \emph{International Conference on Machine Learning}.\hskip 1em plus 0.5em minus 0.4em\relax PMLR, 2023, pp. 26\,106--26\,128.

\bibitem{sun2023don}
Z.~Sun, X.~Du, F.~Song, S.~Wang, M.~Ni, and L.~Li, ``Don't complete it! preventing unhelpful code completion for productive and sustainable neural code completion systems,'' in \emph{2023 IEEE/ACM 45th International Conference on Software Engineering: Companion Proceedings (ICSE-Companion)}.\hskip 1em plus 0.5em minus 0.4em\relax IEEE, 2023, pp. 324--325.

\bibitem{tipirneni2024structcoder}
S.~Tipirneni, M.~Zhu, and C.~K. Reddy, ``Structcoder: Structure-aware transformer for code generation,'' \emph{ACM Transactions on Knowledge Discovery from Data}, vol.~18, no.~3, pp. 1--20, 2024.

\bibitem{ugare2024improving}
S.~Ugare, T.~Suresh, H.~Kang, S.~Misailovic, and G.~Singh, ``Improving llm code generation with grammar augmentation,'' \emph{CoRR}, 2024.

\bibitem{wang2023chatcoder}
Z.~Wang, J.~Li, G.~Li, and Z.~Jin, ``Chatcoder: Chat-based refine requirement improves llms' code generation,'' \emph{arXiv preprint arXiv:2311.00272}, 2023.

\bibitem{yuan2023evaluating}
Z.~Yuan, J.~Liu, Q.~Zi, M.~Liu, X.~Peng, and Y.~Lou, ``Evaluating instruction-tuned large language models on code comprehension and generation,'' \emph{arXiv preprint arXiv:2308.01240}, 2023.

\bibitem{zhu2024hot}
Y.~Zhu, J.~Li, G.~Li, Y.~Zhao, Z.~Jin, and H.~Mei, ``Hot or cold? adaptive temperature sampling for code generation with large language models,'' in \emph{Proceedings of the AAAI Conference on Artificial Intelligence}, vol.~38, no.~1, 2024, pp. 437--445.

\bibitem{li2023starcoder}
R.~Li, L.~B. Allal, Y.~Zi, N.~Muennighoff, D.~Kocetkov, C.~Mou, M.~Marone, C.~Akiki, J.~Li, J.~Chim \emph{et~al.}, ``Starcoder: may the source be with you!'' \emph{arXiv preprint arXiv:2305.06161}, 2023.

\bibitem{finnie2022robots}
J.~Finnie-Ansley, P.~Denny, B.~A. Becker, A.~Luxton-Reilly, and J.~Prather, ``The robots are coming: Exploring the implications of openai codex on introductory programming,'' in \emph{Proceedings of the 24th Australasian computing education conference}, 2022, pp. 10--19.

\bibitem{huang2024effibench}
D.~Huang, Y.~Qing, W.~Shang, H.~Cui, and J.~Zhang, ``Effibench: Benchmarking the efficiency of automatically generated code,'' \emph{Advances in Neural Information Processing Systems}, vol.~37, pp. 11\,506--11\,544, 2024.

\bibitem{liu2024evaluating}
J.~Liu, S.~Xie, J.~Wang, Y.~Wei, Y.~Ding, and L.~Zhang, ``Evaluating language models for efficient code generation,'' \emph{arXiv preprint arXiv:2408.06450}, 2024.

\end{thebibliography}
